\documentclass[aps, prx-quantum, twocolumn, superscriptaddress]{revtex4-2}
\usepackage{amsthm}
\usepackage{amsmath,bm}
\usepackage{amssymb}
\usepackage{amsfonts}
\usepackage{graphicx}
\usepackage{txfonts}
\usepackage{xcolor}
\usepackage{braket}
\usepackage{bbm}
\usepackage{subcaption}
\usepackage{changes}
\usepackage{ragged2e}
\usepackage[colorlinks=true,linkcolor=blue,citecolor=blue,urlcolor=blue]{hyperref}
\usepackage[capitalise]{cleveref}

\newcommand{\ignore}[1]{} 

\newcounter{SaveEqnCntr}

\newcommand{\be}{\begin{equation}}
\newcommand{\ee}{\end{equation}}
\newcommand{\ba}{\begin{eqnarray}}
\newcommand{\ea}{\end{eqnarray}}

\newtheorem{theorem}{Theorem}

\newtheorem{observation}{Observation}

\def\>{\rangle}
\def\<{\langle}

\begin{document}
	
\title{Nonlocality in Continuous-Variable Quantum Networks}	

\author{Sudip Chakrabarty}
\email{sudip27042000@gmail.com}
\affiliation{S. N. Bose National Centre for Basic Sciences, Block JD, Sector III, Salt Lake, Kolkata 700 106, India}

\author{Amit Kundu}
\email{amit$8967$@gmail.com}
\affiliation{S. N. Bose National Centre for Basic Sciences, Block JD, Sector III, Salt Lake, Kolkata 700 106, India}

\author{A. S. Majumdar}
\email{archan@bose.res.in}
\affiliation{S. N. Bose National Centre for Basic Sciences, Block JD, Sector III, Salt Lake, Kolkata 700 106, India}

\keywords{network nonlocality, continuous variables, pseudospin measurement, bilocality, quantum networks}

\begin{abstract}
Quantum networks enable forms of nonlocality beyond the standard Bell scenario, with a multitude of potential applications. Continuous-variable (CV) platforms are particularly attractive for large-scale networks, offering deterministic entanglement generation and favorable prospects for long-distance distribution. Here we present a formalism to study CV network nonlocality using pseudospin measurements. Considering the linear chain and star configurations,  we derive the maximal violations of the corresponding network locality inequalities for arbitrary two-mode states. Using two-mode squeezed vacuum states, we show that the strength of nonlocality in the star configuration remains independent of the network size. Moreover, the nonlocal correlations persist even at arbitrarily high temperatures provided the squeezing exceeds a critical threshold. Further, we demonstrate
non-Gaussianity as an enhancer of network nonlocality through illustrations of various classes of non-Gaussian resources. Remarkably, a coherent superposition of single-photon subtractions across modes achieves maximal violation for vanishing squeezing. Finally, we provide schematics of an experimentally feasible implementation of CV network nonlocality based on the isomorphism between pseudospin and spatial parity observables.
\end{abstract}

\maketitle

\section{Introduction} \label{s1}

Quantum nonlocality, manifested through violations of Bell inequalities~\cite{Bellpaper}, contradicts the classical notion of local realism~\cite{epr} and lies at the heart of quantum information science~\cite{Rev2}. Beyond its foundational significance, nonlocality enables practical tasks such as device-independent protocols~\cite{mckague, mckague12, supicrev, goswami2018, Bian2020}, quantum cryptography~\cite{acin07, Farkas2024, QCrev, Gras2021}, communication complexity~\cite{Rev1}, and randomness certification~\cite{Piro2010, Acin2012}. Rapid experimental progress has turned nonlocality into a reliable resource, with loophole-free demonstrations achieved on multiple platforms, including long-distance optical implementations on telecom networks~\cite{Shalm2015, giustina2013, Hensen2015, Zhong2019, Shin2019, 3gisin17, 4gisin17}. Alongside these developments, nonlocality in continuous-variable (CV) systems remains as an active area of research~\cite{Braunstein2005, Adesso2014, adhikari2008, adhikari2009, chowdhury2013, chowdhury2014}. CV systems are especially attractive due to their natural implementation in quantum optics~\cite{agarwal_book, gerry_book,leibfried1996experimental, smithey1993measurement}. 

Gaussian states, characterized by positive Wigner functions~\cite{Wigner}, are fundamental resources in CV quantum information~\cite{Adesso2014}.
Bell inequality violations for the two-mode squeezed vacuum (TMSV) state
was first demonstrated  using displaced parity measurements~\cite{Banaszek1998}.  A major advance followed with the introduction of pseudospin observables, which enabled maximal Bell violations for Einstein-Podolsky-Rosen (EPR) states and provided an effective framework for probing nonlocality in CV systems~\cite{Chen_Maximal_Bell}. Experimental demonstrations based on spatial parity measurements have further confirmed the nonlocal nature of Gaussian states~\cite{parity_expt}. The established isomorphism between spatial parity and pseudospin observables provides further motivation for exploring this approach. 

On the other hand, the well-known limitations of Gaussian states and operations for tasks such as entanglement distillation and quantum error correction~\cite{Rains1999a, Rains1999b, Bennett1996, Giedke2002, Eisert2002, Braunstein1998} have motivated increasing interest in non-Gaussian states.
Examples of such non-Gaussian states such as those  generated via photon addition or subtraction,  have been realized experimentally and enable
advanced quantum applications through the help of features like enhanced entanglement, stronger Bell violations and EPR steering~\cite{Nongaussian_Waleschaers, Navararrete2012, lee2011, photon_subtracted, Zavatta2004, Parigi2007, Morin2014, boson_sampling, chowdhury2015, maity2017}. These developments naturally motivate elevation of powerful CV nonlocal characteristics of Gaussian and non-Gaussian states   beyond bipartite settings to  the framework of quantum networks.

Quantum networks have emerged as a foundational setting for exploring nonlocal correlations that go beyond those captured by standard Bell-type experiments~\cite{Tavakoli2022}. Unlike standard Bell scenarios with a single source, such networks feature multiple independent entangled sources distributing quantum states to nodes capable of performing joint measurements, allowing richer and stronger correlations than in bipartite settings~\cite{Branciard_2010, Barnciard_2012, gisin_pure, gupta2018, Bithree, kunduBi, Tavakoli_star, Tavakoli2022}. Quantum networks can be used to generate long-distance entanglement distributions without direct interaction between endpoints using entanglement-swapping protocols essential for quantum repeaters~\cite{swap, repeater, repeater2}, and form a fundamental backbone of scalable quantum technologies and the quantum internet~\cite{Kimble2008, repeater, repeater2}. CV systems offer certain key advantages in quantum communication, including deterministic entanglement generation, and suitability for long-range distribution~\cite{Braunstein2005}. Despite such favorable prospects, and a background of a wide range of studies in discrete variable (DV) quantum networks~\cite{Tavakoli2022},  quantum nonlocality in CV networks remain largely unexplored, with only a very limited number of studies in the literature~\cite{GuerraCopete2021, Jiang2025}.

In this work, we investigate nonlocality in quantum networks, focusing on two types of well-known open networks, {\it viz.}, linear chain~\cite{Bithree} and star networks~\cite{Tavakoli_star} involving CV states. We provide a 
formalism for probing nonlocal features in these n-partite CV systems by employing pseudospin measurements. We derive explicit expressions for the maximal violation of the  n-local inequalities that can be achieved by arbitrary two-mode CV states for both
the above types of networks. As our first example, we consider the Gaussian
TMSV state and show that it displays nonlocality in both the chain and star networks  for any nonzero squeezing, with the violation of the corresponding
n-local inequality increasing monotonically and reaching its maximum in the infinite-squeezing limit. Moreover, for the star network with identically
distributed states, we find that the maximal n-locality violation is
independent of the network size.

We extend further the formalism developed here for studying quantum nonlocality in CV networks in several directions. We investigate the robustness of network nonlocality under local thermal noise which, in general, degrades the nonlocal correlations. As a remarkable off-shoot of 
our study, it is observed  that when the squeezing exceeds a critical  value, the state can exhibit nonlocality for any arbitrarily high temperature. 
Moreover, we apply our formalism to examples of popular classes of
non-Gaussian states such as the single photon subtracted TMSV state, the entangled coherent state and the CV Werner state. In addition to subtle
features brought about by the interplay of mode operations, entanglement
 and noise mixing, our results illustrate
enhanced violation of network locality by incorporating non-Gaussianity
compared to the magnitude of nonlocality obtained through the Gaussian
TMSV state. We finally present a strategy for experimentally observing CV network nonlocality using spatial parity observables, by exploiting the well-established isomorphism between the single mode multi-photon electromagnetic field space spanned by the Fock basis in pseudospin space and the single photon multi-mode electromagnetic field space spanned by the spatial eigenmodes~\cite{spatial_parity, parity_expt, Yarnall2007, Kagalwala2017}.

The remainder of the paper is organized as follows. In Sec.~\ref{s2}, we provide a concise overview of the bilocal network scenario and the pseudospin framework for CV systems. Sec.~\ref{s3} presents the general methodology used throughout this work, including the adaptation of DV formalism to CV systems and the strategy for optimizing the associated inequalities focusing on linear chain and star topologies. In Sec.~\ref{s4}, we probe network nonlocality using the TMSV states, followed by an examination of how local thermal noise affects its performance. Sec.~\ref{s5} is devoted to studying
non-Gaussian states. We highlight the impact of single photon subtraction on TMSV states across different subtraction schemes, revealing how such operations can enhance nonlocality, along with displaying similar 
enhancement of nonlocality through the non-Gaussianity of entangled coherent
and CV Werner states. In Sec.~\ref{s6}, we provide an experimental proposal to reveal CV network nonlocality in the spatial parity space. Finally, in Sec.~\ref{s7}, we summarize our main results and suggest potential directions for future research. 

\section{Preliminaries}\label{s2}

This section outlines the essential background for our study. We first recall the bilocal network, which is the simplest example of both linear chain and star network configurations. This is followed by a discussion of pseudospin observables and their relevant properties, which play a central role in our analysis of network nonlocality with CV states.

\subsection{Bilocal versus nonbilocal correlations} \label{s2A}

\begin{figure}[h]
\includegraphics[width=8.5cm]{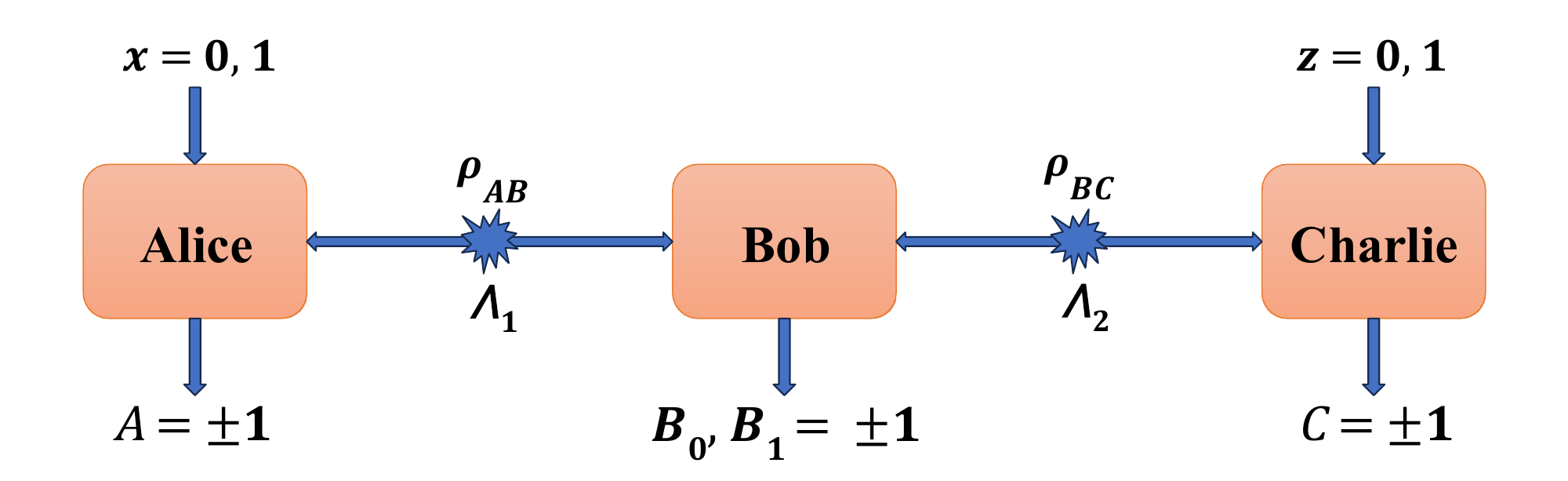}
\caption{Bilocal network scenario}
\label{fig:bilocal1}
\end{figure}

The bilocal network involves three spatially separated observers: Alice, Bob, and Charlie. The central party, Bob, shares an entangled state with Alice, distributed by a source $S_1$, and another entangled state with Charlie, distributed by a second independent source $S_2$ (see Fig.~\ref{fig:bilocal1}).

The defining feature of the bilocal scenario is the independence of the two sources. In a classical hidden variable framework, this implies that the variables associated with $S_1$ and $S_2$, denoted by $\Lambda_1$ and $\Lambda_2$, are statistically independent. Any correlation that respects the bilocal structure must admit a probability decomposition of the form:
\begin{align*}
    p(A,B,C &|x,y,z) = \int \! \! \int d\Lambda_{1} d\Lambda_{2}q_1(\Lambda_{1})q_2(\Lambda_{2}) \\
    &\times \; p(A|x, \Lambda_{1})p(B|y,\Lambda_{1}, \Lambda_{2})p(C|z, \Lambda_{2}),
\end{align*}
where $x$, $y$, and $z$ denote the measurement settings chosen by Alice, Bob, and Charlie, respectively, and $A$, $B$, and $C$ are their corresponding outcomes.

In the quantum realization, Bob receives two particles, one from each entangled pair, and performs a joint measurement. Alice and Charlie each receive one particle from their respective pairs and perform local measurements with binary inputs $x, z \in \{0,1\}$ and outputs $A_x, C_z \in \{\pm1\}$. Bob's measurement yields two binary outcomes, denoted as $B_0, B_1 \in \{\pm 1\}$.

Taking into account the bilocal decomposition, it can be shown that any bilocal hidden variable model must satisfy the following inequality~\cite{gisin_pure}:
\begin{equation}
S_{\mathrm{biloc}} = \sqrt{|I|} + \sqrt{|J|} \leq 2,
\label{bilocal}
\end{equation}
where the correlators $I$ and $J$ are defined as
\begin{align}
I &= \langle(A_{0} + A_{1}) B_{0} (C_{0} + C_{1})\rangle \nonumber\\
&= \langle \sum_{x,z=0}^{1}  A_x B_0 C_z \rangle, \label{IJ1} \\
J &= \langle(A_{0} - A_{1}) B_{1} (C_{0} - C_{1})\rangle \nonumber\\
&= \langle \sum_{x,z=0}^{1} (-1)^{x+z} A_x B_1 C_z \rangle.
\label{IJ2}
\end{align}
Here, $\langle \cdot \rangle$ denotes the expectation value taken over many experimental runs.

Violation of the inequality in Eq.~\eqref{bilocal} provides evidence of nonbilocal correlations, that is, correlations that cannot be explained by any bilocal hidden variable model. This offers a refined notion of nonclassicality, going beyond standard Bell nonlocality by incorporating assumptions about source independence.

\subsection{Pseudospin measurements} \label{s2B}

Pseudospin measurements were introduced to study Bell nonlocality with the TMSV state~\cite{Chen_Maximal_Bell}. Later generalization has enabled one to capture nonlocal correlations using states with odd  photon number difference between the two modes~\cite{zhang2011}. For a single mode light field, the generalized pseudospin operators are defined in the Fock basis $\{\ket{n}\}$ as follows:

\begin{align}
s_x &= \sum_{n=0}^{\infty} \left( \ket{2n+q}\bra{2n+q+1} + \ket{2n+q+1}\bra{2n+q} \right), \nonumber \\
s_y &= i \sum_{n=0}^{\infty} \left( \ket{2n+q}\bra{2n+q+1} - \ket{2n+q+1}\bra{2n+q} \right), \nonumber \\
s_z &= \sum_{n=0}^{\infty} \left( \ket{2n+q+1}\bra{2n+q+1} - \ket{2n+q}\bra{2n+q} \right),  \label{sxyz}
\end{align}
where $q$ is an integer. These operators satisfy the standard $\mathfrak{su}(2)$ algebra, making them formally analogous to Pauli matrices $\boldsymbol{\sigma}^j$, and are thus interpreted as effective spin-$1/2$ operators acting in the photon-number parity space.

The operator $s_z$ corresponds to the negative of the parity operator $P = (-1)^N$, where $N$ is the photon number operator, i.e., $s_z = -P$. The raising and lowering operators $s_{\pm} = \frac{1}{2}(s_x \pm i s_y)$ serve as "parity-flip" operators. In terms of annihilation and creation operators, one can express $s_-$ as~\cite{Chen_Maximal_Bell}:
\[
s_{-} = \frac{I + (-1)^N}{2\sqrt{N+1}} \hat{a},
\]
where $\hat{a}$ is the annihilation operator. The commutation relations among the pseudospin operators are:
\begin{equation}
\left[ s_z, s_{\pm} \right] = \pm 2s_{\pm}, \quad \left[ s_+, s_- \right] = s_z, \label{comm}
\end{equation}
which mirror those of the standard spin-$1/2$ algebra. Hence, the vector operator $\vec{\textbf{s}} = (s_x, s_y, s_z)$ behaves analogously to the spin operator $\vec{\boldsymbol{\sigma}}$, and can be viewed as generating rotations in the "parity spin" space of photons.
In particular, rotations in this space can be described by the unitary operator
\begin{align}
U ({\tau} ,\bf {\hat n}) 
&= \exp \left( -i\frac{\tau }{2}\,{\bf \hat{n}} \cdot \vec{\textbf{s}} \right) \nonumber \\
&= \cos \left( \frac{\tau}{2} \right) - i \left( {\bf \hat{n}} \cdot \vec{\textbf{s}} \right) \sin \left( \frac{\tau}{2} \right), \label{rotate}
\end{align}
where $\tau$ is a rotation angle and ${\bf \hat{n}}$ is an arbitrary unit vector. 

\section{Framework}\label{s3}

As explained in the previous section, the use of pseudospin operators allows for an effective connection between the infinite-dimensional Hilbert space of CV systems and the well-known two-dimensional formalism of spin-1/2 systems. This analogy facilitates the construction of Bell-type measurements and network local inequalities suited for CV states. We begin by defining parity-based Bell like states and outlining partial Bell measurements, followed by the maximization strategy of the network locality inequalities for linear chain and star networks.

\subsection{Partial Bell measurements} \label{s3A}

Following Ref.~\cite{GHZ_cv}, we consider the parity eigenstates associated with mode $j$ of the light field, defined as
\begin{equation}
\left| +\right\rangle_j \equiv \sum_{n=0}^\infty {\cal A}_{n}^{(j)} \left| 2n \right\rangle_j, \quad
\left| -\right\rangle_j \equiv \sum_{n=0}^\infty {\cal B}_{n}^{(j)} \left| 2n+1 \right\rangle_j, \label{par-state}
\end{equation}
where the coefficients ${\cal A}_n^{(j)}$ satisfy the normalization condition $\sum_{n=0}^\infty |{\cal A}_n^{(j)}|^2 = 1$, same for ${\cal B}_{n}^{(j)}$. These parity-defined superpositions of Fock states form a qubit-like basis within an infinite-dimensional Hilbert space.
The action of the pseudospin operators on these states is given by:
\begin{equation}
s_{jz} \ket{\pm}_j = \pm \ket{\pm}_j, \;
s_{j\pm} \ket{\pm}_j = 0, \;
s_{j\pm} \ket{\mp}_j = \ket{\pm}_j. \label{prop}
\end{equation}
Thus, $\ket{\pm}_j$ are eigenstates of $s_{jz}$ with eigenvalues $\pm 1$, and $s_{j\pm}$ serve as raising and lowering operators in the parity space, mirroring the behavior of spin ladder operators in two-level systems.

Using the rotation operator defined in Eq.~(\ref{rotate}), one can construct eigenstates of arbitrary spin projections ${\bf \hat{n}}_j \cdot \vec{\textbf{s}}_j$ for each mode. With this structure in place, one can define analogues of the Bell states for CV systems:
\begin{align}
\ket{\Phi^{+}} &= \frac{1}{\sqrt{2}} \left( \ket{+}_1 \ket{+}_2 + \ket{-}_1 \ket{-}_2 \right), \nonumber \\
\ket{\Phi^{-}} &= \frac{1}{\sqrt{2}} \left( \ket{+}_1 \ket{+}_2 - \ket{-}_1 \ket{-}_2 \right), \nonumber \\
\ket{\Psi^{+}} &= \frac{1}{\sqrt{2}} \left( \ket{+}_1 \ket{-}_2 + \ket{-}_1 \ket{+}_2 \right), \nonumber \\
\ket{\Psi^{-}} &= \frac{1}{\sqrt{2}} \left( \ket{+}_1 \ket{-}_2 - \ket{-}_1 \ket{+}_2 \right). \label{bell_states}
\end{align}
These entangled states are defined entirely within an infinite-dimensional Fock space but retain the algebraic structure of qubit Bell states due to the pseudospin formalism.

We now define the partial Bell measurements performed by Bob using projective measurements in the pseudospin basis. Specifically, the measurement operators are:
\begin{align}
B_0 &= (+1) \times(\ket{\Phi^+}\bra{\Phi^+} + \ket{\Phi^-}\bra{\Phi^-}) \nonumber \\
&+ (-1)\times( \ket{\Psi^+}\bra{\Psi^+} + \ket{\Psi^-}\bra{\Psi^-}) = s_z \otimes s_z, \label{BSM0} \\
B_1 &= (+1) \times(\ket{\Phi^+}\bra{\Phi^+} + \ket{\Psi^+}\bra{\Psi^+}) \nonumber\\
&+ (-1) \times(\ket{\Phi^-}\bra{\Phi^-} + \ket{\Psi^-}\bra{\Psi^-}) = s_x \otimes s_x. \label{BSM1}
\end{align} 
Importantly, $B_0$ and $B_1$ are separable operators composed of local pseudospin measurements. While they distinguish between certain pairs of Bell states, they do not implement a full Bell-state measurement, but rather a partial Bell measurement~\cite{Barnciard_2012}.

\subsection{Bilocal correlations with pseudospin observables} \label{s3B}
Let us choose an arbitrary vector living on the surface of a unit sphere $\boldsymbol{\vec{a}} = (\text{sin} \theta_a \text{cos} \phi_a , \hspace{1mm} \text{sin}\theta_a \text{sin}\phi_a , \hspace{1mm}\text{cos}\theta_a)$, where $\theta_a, \phi_a$ are the polar and azimuthal angles. $\boldsymbol{\vec{a}}$ may be defined as the direction along which we measure parity spin $\boldsymbol{\vec{s}}$. With $\boldsymbol{\vec{s}} = (s_x,s_y,s_z)$, we have,
\begin{equation} \label{a.s}
    \boldsymbol{\vec{a}} \cdot \boldsymbol{\vec{s}} = s_z \cos\theta_a + \sin\theta_a (e^{i\phi_a}s_- +e^{-i\phi_a}s_+).
\end{equation}
Using the commutation relations in Eq.~(\ref{comm}), we have $(\boldsymbol{\vec{a}} \cdot \boldsymbol{\vec{s}})^2 = \boldsymbol{{I}}$, which means that the outcome of the measurements of the Hermitian operator $\boldsymbol{\vec{a}} \cdot \boldsymbol{\vec{s}}$ is $\pm 1$.
Thus, the bilocal inequality in Eq.~(\ref{bilocal}) can be directly used in CV systems considering pseudospin measurements.

To simplify the analysis, we fix $\phi_a = \phi_c =0$ for all the measurements performed by Alice and Charlie, thereby restricting all projective measurements by Alice and Charlie to the X-Z plane. Each measurement of Alice and Charlie can be characterized by one angle. The optimal settings are given by angles symmetric with respect to Z axis, $\pm \theta_1$ for Alice, $\pm \theta_2$ for Charlie. For all $x,z = 0, 1$, we get
\begin{align}
    \langle A_x B_0 C_z\rangle &= \langle [ \cos(\theta_1) s_z +(-1)^x \sin(\theta_1)s_x] \otimes (s_z\otimes s_z) \nonumber
    \\&\otimes [ \cos(\theta_2) s_z +(-1)^z \sin(\theta_2)s_x]\rangle_{\rho_{AB}\otimes \rho_{BC}}\label{I}\\
    \langle A_x B_1 C_z\rangle &= \langle [ \cos(\theta_1) s_z +(-1)^x \sin(\theta_1)s_x] \otimes (s_x\otimes s_x) \nonumber \\
    & \otimes [ \cos(\theta_2) s_z +(-1)^z \sin(\theta_2)s_x]\rangle_{\rho_{AB}\otimes \rho_{BC}} \label{J}
\end{align}
using Eq.~(\ref{IJ1}, \ref{IJ2}), the bilocal expression can be written as 
\begin{align}
    S (\theta_1,\theta_2) = \sqrt{|I(\theta_1,\theta_2)|} + \sqrt{|J(\theta_1,\theta_2)|}  \;,
    \label{bilocal_theta}
\end{align}
with respect to the settings specified by $\theta_1, \theta_2$ for a given state ${\rho_{AB}\otimes \rho_{BC}}$. Our goal is to maximize this expression over the measurement angles. The maximum value is given by, 
\begin{equation}
    S^{\max} = \max_{\theta_1,\theta_2}\hspace{2mm} S(\theta_1,\theta_2)
\end{equation}
Note that the $q$-parameters associated with the pseudospin operators may be chosen by examining the structure of the state. Once the $q$ values  are determined, the remaining task is to optimize over the measurement angles $(\theta_1, \theta_2)$. 

\subsection{Linear Chain Network Scenario} \label{s3C}

\begin{figure}[h]
\includegraphics[width=8.5cm]{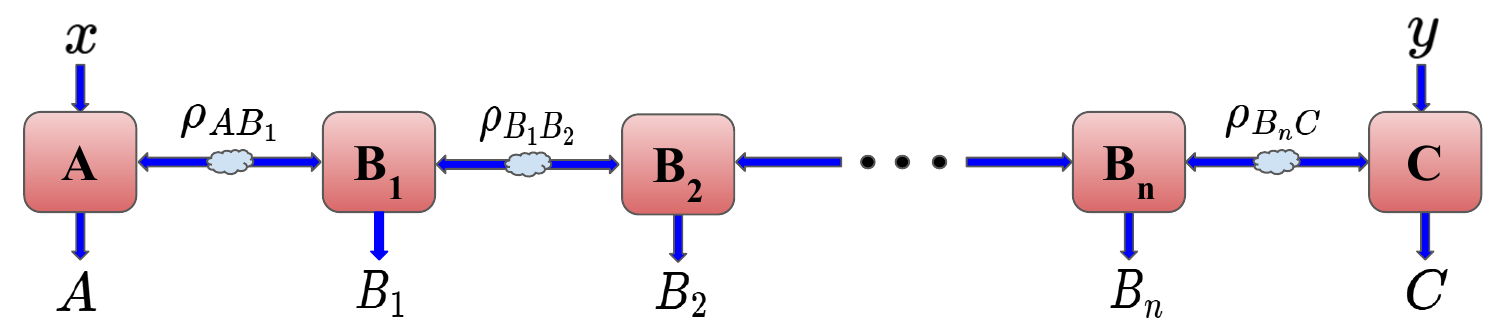}
\caption{Linear Chain network scenario}
\label{fig:linear_chain}
\end{figure}

We now consider a CV quantum network arranged in a linear chain
configuration consisting of $n$ independent sources. The extreme parties,
Alice ($A$) and Charlie ($C$), are connected through a sequence of intermediate
nodes $B_1,B_2,\dots,B_{n-1}$ (see Fig.~\ref{fig:linear_chain}). Each source distributes a two mode state $\rho_{B_k B_{k+1}}$, with the identification $B_0 \equiv A$ and
$B_{n} \equiv C$. The global state of the network is therefore
\begin{equation}
\rho = \bigotimes_{k=0}^{n-1} \rho_{B_k B_{k+1}} .
\end{equation}

Alice and Charlie each perform one of two dichotomic pseudospin measurements,
specified by the setting pairs
$(\boldsymbol{\vec{a}},\boldsymbol{\vec{a}'})$ and
$(\boldsymbol{\vec{c}},\boldsymbol{\vec{c}'})$, respectively.
Each intermediate node $B_k$ performs a fixed joint pseudospin measurement on
the two modes received from the adjacent sources. These measurements are chosen
such that, along one branch of the inequality, the correlations involve
pseudospin $s_z$ operators, while along the complementary branch they involve
$s_x$ operators.

Under the assumption of $n$-local hidden variables
$\Lambda_1,\dots,\Lambda_{n}$, the observable correlations must satisfy the
nonlinear $n$-local inequality~\cite{Bithree},
\begin{equation}
S_{\mathrm{chain}}
=
\sqrt{|I_n^{\mathrm{chain}}|}
+
\sqrt{|J_n^{\mathrm{chain}}|}
\le 2 .
\label{eq:nlocal}
\end{equation}
The correlators are defined as
\begin{align}
I_n^{\mathrm{chain}} &=
\Big\langle
(\boldsymbol{\vec{a}}+\boldsymbol{\vec{a}'})\!\cdot\!\boldsymbol{\vec{s}}
\otimes (s_z\otimes s_z)^{\otimes {(n-1)}}
\otimes (\boldsymbol{\vec{c}}+\boldsymbol{\vec{c}'})\!\cdot\!\boldsymbol{\vec{s}}
\Big\rangle , \label{eq:Ichain} \\
J_n^{\mathrm{chain}} &=
\Big\langle
(\boldsymbol{\vec{a}}-\boldsymbol{\vec{a}'})\!\cdot\!\boldsymbol{\vec{s}}
\otimes (s_x\otimes s_x)^{\otimes {(n-1)}}
\otimes (\boldsymbol{\vec{c}}-\boldsymbol{\vec{c}'})\!\cdot\!\boldsymbol{\vec{s}}
\Big\rangle . \label{eq:Jchain}
\end{align}

Because the global state is a tensor product over independent sources and all
observables are pseudospin operators, the correlators factorize into products
of two-mode correlations. Introducing the pseudospin correlation matrices
\begin{equation}
t^{(k)}_{ij}
=
\mathrm{Tr}\!\left[ s_i \otimes s_j \, \rho_{B_k B_{k+1}} \right],
\qquad i,j \in \{1,2,3\},
\end{equation}
with $(s_1,s_2,s_3) = (s_x,s_y,s_z)$, Eqs.~(\ref{eq:Ichain}) and
(\ref{eq:Jchain}) reduce to
\begin{align}
I_n^{\mathrm{chain}}
&=\text{Tr}\left[(\boldsymbol{\vec{a} + \vec{a}'}) \cdot \boldsymbol{\vec{s}} \otimes s_z \, \rho_{B_0 B_1} \right] \times \left(\prod_{k=1}^{n-2} \text{Tr}\left[ s_z \otimes s_z\rho_{B_k B_{k+1}}\right] \right) \nonumber\\
& \qquad \times\text{Tr}\left[ s_z \otimes (\boldsymbol{\vec{c} + \vec{c}'}) \cdot \boldsymbol{\vec{s}} \, \rho_{B_{n-1}B_{n}} \right] \nonumber \\
&=
\sum_i (a_i+a_i')\, t^{(0)}_{i3}
\left( \prod_{k=1}^{n-2} t^{(k)}_{33} \right)
\sum_j t^{(n-1)}_{3j}\,(c_j+c_j'), \label{eq:Ifactor} \\
J_n^{\mathrm{chain}}
&=
\text{Tr}\left[(\boldsymbol{\vec{a} - \vec{a}'}) \cdot \boldsymbol{\vec{s}} \otimes s_x \, \rho_{B_0 B_1} \right] \times \left(\prod_{k=1}^{n-2} \text{Tr}\left[ s_x \otimes s_x\rho_{B_k B_{k+1}}\right] \right) \nonumber\\
&\qquad \times \text{Tr}\left[ s_x \otimes (\boldsymbol{\vec{c} - \vec{c}'}) \cdot \boldsymbol{\vec{s}} \, \rho_{B_{n-1}B_{n}} \right] \nonumber \\
&=
\sum_i (a_i-a_i')\, t^{(0)}_{i1} 
\left( \prod_{k=1}^{n-2} t^{(k)}_{11} \right)
\sum_j t^{(n-1)}_{1j}\,(c_j-c_j'). \label{eq:Jfactor}
\end{align}

Now we derive a closed-form expression for the maximal value of the n-locality expression achievable by arbitrary two mode CV states, under the restriction that all local measurements are pseudospin observables. This allows one to quantify network nonlocal correlations of CV states using a finite set of effective pseudospin measurements, and provides a direct analog of the well-known Horodecki criterion for Bell nonlocality in bipartite systems~\cite{Horodecki_Bell}. Our approach generalizes the correlation-matrix optimization method originally developed for two-qubit states~\cite{gisin_pure} and later extended to $n$-local network scenarios~\cite{kunduBi, Andreoli_2017}.

\begin{theorem}
For a linear chain network probed by pseudospin measurements, the maximal value
of the $n$-locality expression is
\begin{equation}
S_{\mathrm{chain}}^{\max}
=
2
\sqrt{
\left(\prod_{k=0}^{n-1} \nu_1^{(k)}\right)
+
\left(\prod_{k=0}^{n-1} \nu_2^{(k)}\right)
},
\label{eq:Schainmax}
\end{equation}
where $\nu_1^{(k)} \ge \nu_2^{(k)} \ge \nu_3^{(k)} \ge 0$ are the singular values of
the correlation matrix $t^{(k)}$ associated with the state
$\rho_{B_k B_{k+1}}$. Consequently, the $n$-local inequality
(\ref{eq:nlocal}) is violated whenever
\begin{equation}
\prod_{k=0}^{n} \nu_1^{(k)} + \prod_{k=0}^{n} \nu_2^{(k)} > 1 .
\label{eq:violation}
\end{equation}
\end{theorem}

\begin{proof}
For each source $k$, we perform a singular-value decomposition of the correlation
matrix $t^{(k)}$, with $R^{(k)} = \sqrt{(t^{(k)})^\dagger t^{(k)}}$.
The non-negative eigenvalues of $R^{(k)}$ are denoted as,
$\nu_1^{(k)} \ge \nu_2^{(k)} \ge \nu_3^{(k)}$.

The extremal parties Alice and Charlie can restrict their measurement directions
to the two-dimensional subspaces spanned by the singular vectors corresponding
to $\nu_1^{(0)},\nu_2^{(0)}$ and $\nu_1^{(n-1)},\nu_2^{(n-1)}$, respectively, without
loss of generality. This simplification allows the use of the following parameterization:
\begin{align*}
    \boldsymbol{\vec{a}} &= (\sin\theta_1, 0, \cos\theta_1),\\
    \boldsymbol{\vec{a}'}&= (\sin\theta_1', 0, \cos\theta_1') \\
    \boldsymbol{\vec{c}} &= (\sin\theta_2, 0, \cos\theta_2),\\
    \boldsymbol{\vec{c}'}&= (\sin\theta_2', 0, \cos\theta_2')
\end{align*}

Under these choices, Eqs.~(\ref{eq:Ifactor}) and (\ref{eq:Jfactor}) reduce to
effective two-dimensional expressions,
\begin{align}
I_n^{\mathrm{chain}} &=
\left(\prod_{k=0}^{n-1} \nu_1^{(k)}\right)
(\cos\theta_1 + \cos\theta_1')
(\cos\theta_2 + \cos\theta_2'), \\
J_n^{\mathrm{chain}} &=
\left(\prod_{k=0}^{n-1} \nu_2^{(k)}\right)
(\sin\theta_1 - \sin\theta_1')
(\sin\theta_2 - \sin\theta_2').
\end{align}

The optimal angles are obtained by maximizing the n-locality expression with respect to the measurement parameters. Setting the partial derivatives of $S_{\text{chain}}$ with respect to all angles to zero yields the following optimality conditions:
\[
\theta_1' = -\theta_1, \qquad \theta_2' = -\theta_2,
\]
together with
\[
\cos\theta_1 = \cos\theta_2 =
\sqrt{
\frac{\prod_{k=0}^{n-1} \nu_1^{(k)}}
{\prod_{k=0}^{n-1} \nu_1^{(k)} + \prod_{k=0}^{n-1} \nu_2^{(k)}}
}.
\]
Substituting these values into Eq.~(\ref{eq:nlocal}) yields the maximal value
$S_{\mathrm{chain}}^{\max}$ in Eq.~(\ref{eq:Schainmax}).
\end{proof}

In the symmetric case where all sources distribute identical states,
$t^{(k)} = t$ for all $k$, the violation condition simplifies to
\begin{equation}
(\nu_1)^{n} + (\nu_2)^{n} > 1 ,
\end{equation}
This condition shows that network nonlocality weakens with increasing chain length. Since $\nu_1,\nu_2 \leq 1$ for physical states, the contributions $(\nu_1)^{n}$ and $(\nu_2)^{n}$ either or both decrease monotonically as $n$ grows. Consequently, noisy sources with $\nu_1,\nu_2 < 1$ cease to violate the inequality beyond a certain network size.

\subsection{Star network scenario} \label{s3D}

\begin{figure}[h]
\includegraphics[width=7cm]{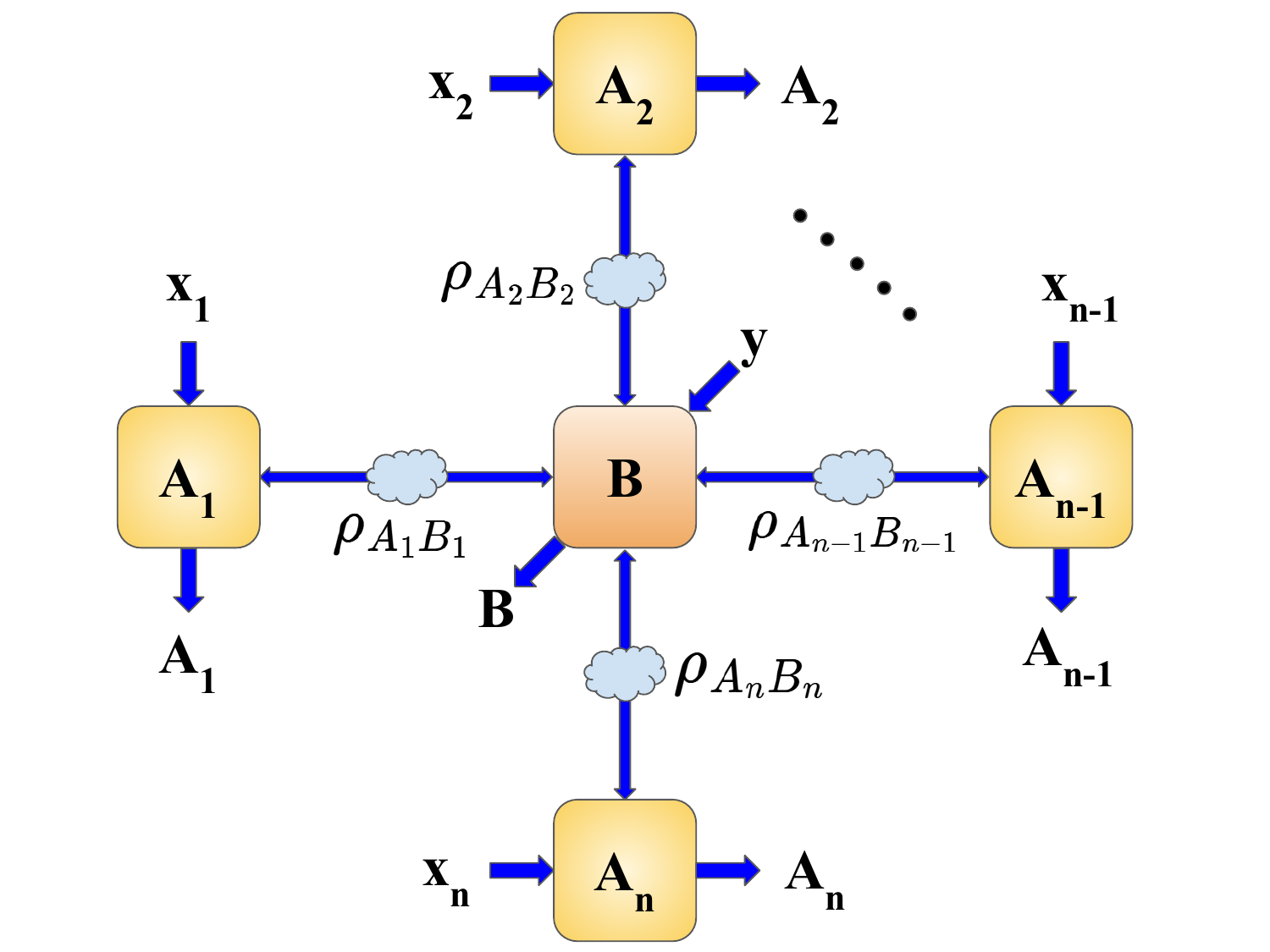}
\caption{Star network scenario}
\label{fig:star}
\end{figure}

We now extend the correlation matrix optimization method to $n$-partite star network \cite{Tavakoli_star, Andreoli_2017, kunduBi} probed by pseudospin
measurements on CV systems. This scenario generalizes the bilocal
configuration and consists of $n$ independent sources, each distributing
a two mode CV state between one peripheral party $A_i$ and a central
node $B$ (see Fig.~\ref{fig:star}).

The global state of the network is assumed to be a tensor product of
independent bipartite states,
\begin{equation}
\rho
=
\bigotimes_{i=1}^{n} \rho_{A_i B_i},
\end{equation}
where the states $\rho_{A_i B_i}$ are arbitrary two mode CV states. Each peripheral
party $A_i$ performs one of two dichotomic pseudospin measurements,
\begin{equation}
A^i_{x_i}
=
\boldsymbol{\vec a}^{\,i}_{x_i}\cdot\boldsymbol{\vec s},
\qquad x_i\in\{0,1\},
\end{equation}
while the central node performs separable joint measurements of the form
\begin{equation}
B_y
=
\bigotimes_{i=1}^{n} B^i_y,
\qquad
B^i_y
=
\boldsymbol{\vec b}^{\,i}_y\cdot\boldsymbol{\vec s},
\qquad y\in\{0,1\},
\end{equation}
with $\boldsymbol{\vec s}=(s_x,s_y,s_z)$ denoting the pseudospin operators.

Under the assumption of $n$-local hidden variables, the local correlations
observed in this network satisfy the nonlinear star-network inequality~\cite{Tavakoli_star}
\begin{equation}
\mathcal{S}_{\mathrm{star}}
=
|I|^{1/n}
+
|J|^{1/n}
\leq 2,
\label{eq:star_ineq}
\end{equation}
where
\begin{align}
I_n^{star} &=
\sum_{x_1,\dots,x_n}
\big\langle
A^1_{x_1}\cdots A^n_{x_n} B_0
\big\rangle,
\\
J_n^{star} &=
\sum_{x_1,\dots,x_n}
(-1)^{\sum_i x_i}
\big\langle
A^1_{x_1}\cdots A^n_{x_n} B_1
\big\rangle .
\end{align}

Because the global state is a tensor product over independent sources and
the measurement at the central node is separable, the correlators
factorize as
\begin{equation}
\big\langle
A^1_{x_1}\cdots A^n_{x_n} B_y
\big\rangle
=
\prod_{i=1}^{n}
\big\langle
A^i_{x_i} B^i_y
\big\rangle .
\end{equation}

For each bipartite state $\rho_{A_i B_i}$ we define the pseudospin
correlation matrix
\begin{equation}
t^{(i)}_{jk}
=
\mathrm{Tr}
\!\left[
s_j\otimes s_k\;
\rho_{A_i B_i}
\right],
\qquad j,k\in\{x,y,z\}.
\end{equation}
Let $R^{(i)}=\sqrt{(t^{(i)})^{T}t^{(i)}}$ and denote its eigenvalues by
$\nu^{(i)}_1\ge\nu^{(i)}_2\ge\nu^{(i)}_3\ge0$. Now we derive a closed-form expression for the maximal value of the n-locality expression achievable by arbitrary two mode CV states, in the star network.

\begin{theorem}
\label{thm:star_cv_max}
For an $n$-partite CV star network probed by pseudospin measurements and
described by the global state
$\rho=\bigotimes_{i=1}^n \rho_{A_iB_i}$, the maximal value of the
star-network expression $\mathcal{S}_{\mathrm{star}}$ defined in
Eq.~\eqref{eq:star_ineq} is given by
\begin{equation}
\mathcal{S}_{\mathrm{star}}^{\max}
=
2\sqrt{
\left(
\prod_{i=1}^{n}\nu^{(i)}_1
\right)^{\!2/n}
+
\left(
\prod_{i=1}^{n}\nu^{(i)}_2
\right)^{\!2/n}
},
\label{eq:star_cv_max}
\end{equation}
where $\nu^{(i)}_1\ge\nu^{(i)}_2\ge\nu^{(i)}_3\ge0$ are the eigenvalues of
$R^{(i)}$. Consequently, the star-network inequality is
violated whenever
\begin{equation}
\left(
\prod_{i=1}^{n}\nu^{(i)}_1
\right)^{\!2/n}
+
\left(
\prod_{i=1}^{n}\nu^{(i)}_2
\right)^{\!2/n}
>1 .
\end{equation}
\end{theorem}

\begin{proof}
Using the pseudospin representation, the two-body correlators entering
$I$ and $J$ take the form
\begin{equation}
\langle A^i_{x_i} B^i_y\rangle
=
\boldsymbol{\vec a}^{\,i}_{x_i}\cdot
t^{(i)}
\boldsymbol{\vec b}^{\,i}_y .
\end{equation}
Because the global state is a tensor product over independent sources and
the measurement at the central node is separable, all correlators
factorize \cite{Tavakoli_star}, yielding
\begin{align}
\mathcal{S}_{\mathrm{star}}
&=
\left|
\prod_{i=1}^n
\big(
\langle A^i_0 B^i_0\rangle
+
\langle A^i_1 B^i_0\rangle
\big)
\right|^{\!1/n} \nonumber \\
&+
\left|
\prod_{i=1}^n
\big(
\langle A^i_0 B^i_1\rangle
-
\langle A^i_1 B^i_1\rangle
\big)
\right|^{\!1/n}.
\end{align}

Introducing the orthogonal decompositions
\begin{align}
\boldsymbol{\vec a}^{\,i}_0 + \boldsymbol{\vec a}^{\,i}_1
& = 2 \cos\theta_i\,\boldsymbol{\vec n}_i \;,\; \nonumber\\
\boldsymbol{\vec a}^{\,i}_0 - \boldsymbol{\vec a}^{\,i}_1
&= 2\sin\theta_i\,\boldsymbol{\vec n}'_i , \nonumber\\
\boldsymbol{\vec n}_i\cdot\boldsymbol{\vec n}'_i &= 0,
\end{align}
and optimizing over the measurement directions at the central node, one
obtains

Substituting the above decomposition into the correlators,
we obtain
\begin{align}
\langle A^i_0 B^i_0\rangle + \langle A^i_1 B^i_0\rangle
&=
(\boldsymbol{\vec a}^{\,i}_0+\boldsymbol{\vec a}^{\,i}_1)
\cdot t^{(i)}\boldsymbol{\vec b}^{\,i}_0 \nonumber\\
&=
2\cos\theta_i\;
\boldsymbol{\vec n}_i\cdot t^{(i)}\boldsymbol{\vec b}^{\,i}_0 ,
\\[2mm]
\langle A^i_0 B^i_1\rangle - \langle A^i_1 B^i_1\rangle
&=
(\boldsymbol{\vec a}^{\,i}_0-\boldsymbol{\vec a}^{\,i}_1)
\cdot t^{(i)}\boldsymbol{\vec b}^{\,i}_1 \nonumber\\
&=
2\sin\theta_i\;
\boldsymbol{\vec n}'_i\cdot t^{(i)}\boldsymbol{\vec b}^{\,i}_1 .
\end{align}

Hence,
\begin{align}
\mathcal{S}_{\mathrm{star}}
&=
2\Bigg[
\left|
\prod_{i=1}^n
\cos\theta_i\;
\boldsymbol{\vec n}_i\cdot t^{(i)}\boldsymbol{\vec b}^{\,i}_0
\right|^{1/n}
\nonumber\\
&\hspace{1.2cm}
+
\left|
\prod_{i=1}^n
\sin\theta_i\;
\boldsymbol{\vec n}'_i\cdot t^{(i)}\boldsymbol{\vec b}^{\,i}_{1}
\right|^{1/n}
\Bigg].
\end{align}

For fixed $\boldsymbol{\vec n}_i$ and $\boldsymbol{\vec n}'_i$, the
maximization over each measurement direction
$\boldsymbol{\vec b}^{\,i}_y$ can be performed independently.
Using
\begin{equation}
\max_{\|\boldsymbol{\vec b}\|=1}
\boldsymbol{\vec u}\cdot t^{(i)}\boldsymbol{\vec b}
=
\|(t^{(i)})^{T}\boldsymbol{\vec u}\|,
\end{equation}
which is achieved when
\(
\boldsymbol{\vec b}^{\,i}_y
\propto (t^{(i)})^{T}\boldsymbol{\vec u},
\)
we obtain Eq.~\eqref{eq:Sstar_intermediate}.

\begin{align}
\mathcal{S}_{\mathrm{star}}
&=
2\Bigg[
\left|
\prod_{i=1}^n
\cos\theta_i\;
\|(t^{(i)})^{T}\boldsymbol{\vec n}_i\|
\right|^{\!1/n}
\nonumber\\
&\hspace{1.2cm}
+
\left|
\prod_{i=1}^n
\sin\theta_i\;
\|(t^{(i)})^{T}\boldsymbol{\vec n}'_i\|
\right|^{\!1/n}
\Bigg].
\label{eq:Sstar_intermediate}
\end{align}

Next, we optimize over the angles $\{\theta_i\}$. Defining
\begin{equation} \label{R_theta}
\mathcal{R}(\theta_1,\dots,\theta_n)
=
\left|
\prod_{i=1}^n \delta_1 \cos\theta_i
\right|^{1/n}
+
\left|
\prod_{i=1}^n \delta_2 \sin\theta_i
\right|^{1/n},
\end{equation}
where
\(
\delta_1= \left|\prod_{i=1}^n \|(t^{(i)})^{T}\boldsymbol{\vec n}_i\|
\, \right| \)
and
\(
\delta_2=\left|\prod_{i=1}^n \|(t^{(i)})^{T}\boldsymbol{\vec n}'_i\|\, \right|
\). From the stationarity conditions
$\partial \mathcal{R}/\partial\theta_j=0$, we get,

\begin{align}
    \frac{\partial \mathcal{R}}{\partial \theta_j} &= - \frac{1}{n}\left[  \tan \theta_j \left|
\delta_1
\prod_{i=1}^n \cos\theta_i
\right|^{1/n}  - \cot \theta_j \left|
\delta_2
\prod_{i=1}^n \sin\theta_i
\right|^{1/n} \right] \nonumber \\
&=0,
\end{align}
which leads to the constraint,
\begin{align} \label{theta_j}
    \tan^2\theta_j=\tan^2\theta_i, \qquad \forall i,j.
\end{align}
We plug this constraint from Eq.~\eqref{theta_j} in Eq.~\eqref{R_theta}. We get,
\begin{align}
\max_{\{\theta_i\}} \mathcal{R}
&= \sqrt{\delta_1^{2/n}+\delta_2^{2/n}} \nonumber\\
&=\sqrt{
\left|
\prod_{i=1}^n
\|(t^{(i)})^{T}\boldsymbol{\vec n}_i\|
\right|^{2/n}
+
\left|
\prod_{i=1}^n
\|(t^{(i)})^{T}\boldsymbol{\vec n}'_i\|
\right|^{2/n}
}.
\end{align}

Substituting this back into Eq.~\eqref{eq:Sstar_intermediate}, we obtain
\begin{equation}
\mathcal{S}_{\mathrm{star}}^{\max}
=
2\max
\sqrt{
\left|
\prod_{i=1}^n
\|(t^{(i)})^{T}\boldsymbol{\vec n}_i\|
\right|^{2/n}
+
\left|
\prod_{i=1}^n
\|(t^{(i)})^{T}\boldsymbol{\vec n}'_i\|
\right|^{2/n}
}.
\label{eq:Sstar_after_alpha}
\end{equation}

Now, we perform the optimization over the directions
$\boldsymbol{\vec n}_1$ and $\boldsymbol{\vec n}'_1$. Fixing all vectors
except those corresponding to the preferred source (here $i=1$), and defining
\begin{equation}
k_1=
\left| \prod_{i=2}^n \|(t^{(i)})^{T}\boldsymbol{\vec n}_i\|\,\right|,
\qquad
k_2=
\left| \prod_{i=2}^n \|(t^{(i)})^{T}\boldsymbol{\vec n}'_i\|\, \right|,
\end{equation}
Eq.~\eqref{eq:Sstar_after_alpha} reduces to
\begin{equation}
\mathcal{S}_{\mathrm{star}}^{\max}
=
2 \max
\sqrt{
k_1^{2/n}\|(t^{(1)})^{T}\boldsymbol{\vec n}_1\|^{2/n}
+
k_2^{2/n}\|(t^{(1)})^{T}\boldsymbol{\vec n}'_1\|^{2/n}
}.
\end{equation}

We recall that $\nu^{(1)}_1\ge\nu^{(1)}_2\ge\nu^{(1)}_3$ denote the eigenvalues of $R^{(1)} \left(=\sqrt{(t^{(1)})^{T}t^{(1)}} \right)$. Writing $\boldsymbol{\vec n}_1$ and
$\boldsymbol{\vec n}'_1$ in the corresponding eigenbasis, the maximization
under the constraints
\(
\|\boldsymbol{\vec n}_1\|=\|\boldsymbol{\vec n}'_1\|=1
\)
and
\(
\boldsymbol{\vec n}_1\cdot\boldsymbol{\vec n}'_1=0
\)
is solved using Lagrange multipliers. 

To see this explicitly,
\begin{equation}
\|(t^{(1)})^{T}\boldsymbol{\vec n}_1\|^2
=
\boldsymbol{\vec n}_1^{T}(t^{(1)})^{T}t^{(1)}\boldsymbol{\vec n}_1
=
\boldsymbol{\vec n}_1^{T}(R^{(1)})^2\boldsymbol{\vec n}_1 .
\end{equation}
Expanding
\(
\boldsymbol{\vec n}_1=\sum_j c_j \boldsymbol{\vec \nu}_j
\)
in the eigenbasis $\{\boldsymbol{\vec \nu}_j\}$ of $R^{(1)}$ gives
\begin{equation}
\|(t^{(1)})^{T}\boldsymbol{\vec n}_1\|^2
=
\sum_j c_j^2 (\nu_j^{(1)})^2 ,
\qquad
\sum_j c_j^2=1.
\end{equation}
This quantity is maximized when $c_1=1$, implying alignment with the
eigenvector corresponding to $\nu^{(1)}_1$.
Orthogonality then forces $\boldsymbol{\vec n}'_1$ to align with the eigenvector corresponding to $\nu^{(1)}_2$. We obtain,
\begin{equation}
\mathcal{S}_{\mathrm{star}}^{\max} = 2\max{
\sqrt{(k_1)^{2/n} (\nu_1^{(1)})^{2/n}  + (k_2)^{2/n} (\nu_2^{(1)})^{2/n}}}.
\end{equation}\label{eq:Sstar_after_theta1}

The same optimization can now be carried out for
$\boldsymbol{\vec n}_2,\boldsymbol{\vec n}'_2$ while keeping all remaining
vectors fixed. Repeating this procedure sequentially for
$i=2,\dots,n$ replaces each pair
\(
\|(t^{(i)})^{T}\boldsymbol{\vec n}_i\|
\)
and
\(
\|(t^{(i)})^{T}\boldsymbol{\vec n}'_i\|
\)
by
$\nu^{(i)}_1$ and $\nu^{(i)}_2$, respectively.
After completing the optimization over all sources,
Eq.~\eqref{eq:Sstar_after_alpha} reduces to
\begin{equation}
\mathcal{S}_{\mathrm{star}}^{\max}
=
2\sqrt{
\left(
\prod_{i=1}^{n}\nu^{(i)}_1
\right)^{2/n}
+
\left(
\prod_{i=1}^{n}\nu^{(i)}_2
\right)^{2/n}
},
\end{equation}
which coincides with Eq.~\eqref{eq:star_cv_max}.
This completes the proof.
\end{proof}

\subsection{Comparison with two qubit systems} For two-qubit systems, violation of network local inequalities in the linear chain and star networks based on Pauli operators offer necessary and sufficient conditions for detecting nonlocality \cite{gisin_pure}. This is due to the fact that Pauli operators span the full operator space. In contrast, pseudospin operators in CV systems form only a subspace of all dichotomic observables. Hence, the pseudospin-based network local inequalities used here provide a sufficient but not necessary condition for nonlocality. Obtaining a necessary condition would require optimization over all dichotomic observables in CV systems, a task that is practically infeasible as the basis for constructing any dichotomic observable is also infinite. Nevertheless, pseudospin observables have proven to be effective and one of the best options in various studies, achieving Tsirelson’s bound for EPR states \cite{Chen_Maximal_Bell} where other methods \cite{Banaszek1998} fall short, and have been successfully extended to broader classes of states \cite{zhang2011, dorantes2009}. 

\section{Network Nonlocality with Two-Mode Squeezed Vacuum States} \label{s4}

Gaussian states play a central role in the analysis of CV quantum systems \cite{Adesso2014}. Within this class, the most general form of pure states is given by displaced squeezed states \cite{Braunstein2005}. However, for the purpose of investigating quantum correlations, it is well known that the displacement operation does not affect a state's nonlocal features~\cite{Braunstein2005}. Therefore, without any loss of generality, we restrict our analysis to TMSV, excluding displacements. This state is obtained by squeezing the two-mode vacuum state ($\ket{0,0}$) using the two-mode squeezing operator $\mathcal{S}(\zeta) = \exp(\zeta \hat{a}_{1}^{\dagger} \hat{a}_{2}^{\dagger} - {\zeta}^{*} \hat{a}_1 \hat{a}_2)$ where $\zeta = r \exp(i\phi)$ and $r > 0$ is the squeezing parameter. This state can be produced in a nondegenerate optical parametric amplifier. Considering $\phi=0$, the state is represented as \cite{Braunstein2005}:
\begin{equation}
\begin{split}
    |\zeta\rangle = \mathcal{S}(\zeta)\ket{0,0} = \sqrt{1-\lambda^2}\sum_{n-0}^{\infty}\lambda^n\ket{n,n} \label{nopa}
\end{split}
\end{equation} 
where, $\lambda = \tanh{r}\in [0,1]$, and $|n,n\rangle = |n\rangle \otimes |n\rangle$ are usual two mode Fock states. 
Pseudospin observables with $q_i =0$ can be chosen to show maximal violation with TMSV \cite{Chen_Maximal_Bell}.

First, we consider the bilocal network scenario where the states shared between Alice and Bob $(\ket{\psi_{AB}})$, and the state shared between Bob and Charlie $(\ket{\psi_{BC}}$ are TMSVs with squeezing parameters $r_1$ and $r_2$. We represent them as $|\zeta\rangle_1$ and $|\zeta\rangle_2$ respectively. Here we consider pseudospin observables with all $q_i=0$. Using Eq.~(\ref{I}) and Eq.~(\ref{J}), we get,
\begin{align}
    \langle A_x B_0 C_z\rangle_{|\zeta\rangle_1 \otimes |\zeta\rangle_2} &=  \cos(\theta_1)\cos(\theta_2)
    \\ \langle A_x B_1 C_z\rangle_{|\zeta\rangle_1 \otimes |\zeta\rangle_2} &=  K(r_1)K(r_2) \sin(\theta_1) \sin(\theta_2)
\end{align}
where $K(r_1) = \text{tanh} (2r_1),\hspace{2mm} K(r_2 )= \text{tanh}(2r_2)$.
Using Eq.~(\ref{IJ1}), and Eq.~ (\ref{IJ2}), we can calculate the values of $I$, $J$ and $S$,
\begin{align}
    I_{\zeta_1, \zeta_2} &= 4 \cos(\theta_1)\cos(\theta_2) \nonumber\\
    J_{\zeta_1, \zeta_2} &=  4K(r_1)K(r_2) \sin(\theta_1) \sin(\theta_2) \nonumber\\
       &= 4L^2 \sin(\theta_1) \sin(\theta_2) \nonumber\\
    S_{\zeta_1, \zeta_2} &= 2 \left[ \sqrt{\cos(\theta_1)\cos(\theta_2)} +L \sqrt{\sin(\theta_1) \sin(\theta_2)}  \right] 
\end{align}
where, $L^2 = K(r_1)K(r_2)$. 
The optimal measurement settings, which maximizes the value of $S_{\zeta}$ in Eq.~(\ref{bilocal_theta}), is given by,
\begin{align}
& \theta_1 = \theta_2 = \theta = \arctan (L)\\
 \text{Hence,}\hspace{2mm}  \cos & \theta = \frac{1}{\sqrt{1+L^2}} \hspace{2mm} \text{and} \hspace{2mm} \sin{\theta} = \frac{L}{\sqrt{1+L^2}} \nonumber
\end{align}
such that the maximal bilocality quantity $S^{\max}$ reduces to,
\begin{align}
    S^{\max}_{\zeta_1, \zeta_2} = 2\sqrt{1+L^2}
\end{align}
Thus, the bilocality inequality Eq.~(\ref{bilocal}) is always violated by TMSV provided $r_1, r_2 \neq 0$, and the degree of violation is completely specified by the squeezing parameters $r_1, r_2$. 
In the infinite squeezing limit, $r_1,r_2\rightarrow \infty$, $K(r_1), K(r_2)\rightarrow 1$, both $\ket{\zeta_1}$ and $\ket{\zeta_2}$ become the original EPR states, for which we have,
\begin{align}
    S^{\max}_{\zeta_1, \zeta_2} (r_1 \rightarrow\infty, r_2 \rightarrow \infty)= 2\sqrt{2}
\end{align}
This indicates that the normalized version of original EPR states can maximally violate the bilocality inequality (see Fig.~\ref{fig:star_and_chain_comparison}).

\subsection{Linear-chain and star networks with TMSV states}
\label{s4A}

We now investigate $n$-local correlations in the linear chain and the star networks, assuming that each source distributes a TMSV state with identical squeezing parameter $r$. For each bipartite source, the correlations are characterized by the correlation matrix $t^{(i)}$ with elements
\begin{equation}
t^{(i)}_{jk}
=
\mathrm{Tr}
\!\left[
s_j \otimes s_k \;
\ket{\zeta}\bra{\zeta}
\right],
\qquad j,k \in \{x,y,z\},
\end{equation}
where $\{s_x, s_y, s_z\}$ denote the pseudospin operators.

The nonzero eigenvalues of the associated matrix
\(
R^{(i)} = \sqrt{(t^{(i)})^{T} t^{(i)}}
\)
are found to be
\begin{align}
\mu_1 = 1, \qquad \mu_2 = K(r) = \tanh(2r),
\end{align}
which completely determine the maximal violation of the $n$-local network inequalities following the prescription introduced in Sec.~\ref{s3}.

\begin{figure}
    \centering
    \includegraphics[width=8cm]{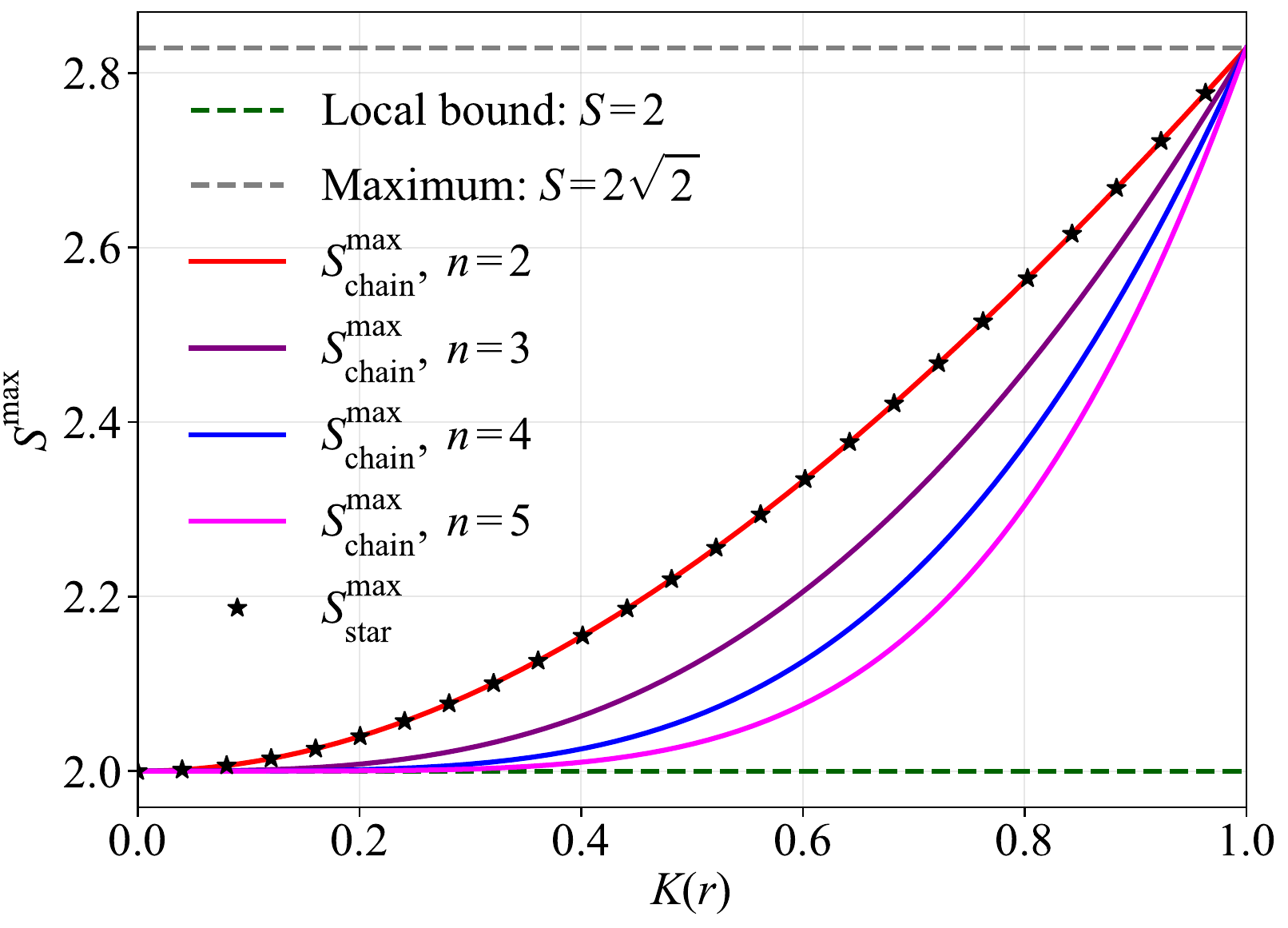}
\caption{
\justifying
Maximal $n$-locality expressions for linear-chain and star network configurations using TMSV states as a function of the squeezing parameter $K = \tanh(2r)$, assuming equal squeezing $r_i = r$ for all sources. The violations increase monotonically with squeezing and approach the maximal quantum value $2\sqrt{2}$ in the limit $K \rightarrow 1$. The horizontal dashed lines at $S = 2$ (green) and $S = 2\sqrt{2}$ (gray) denote the classical and maximal quantum bounds, respectively.
}
    \label{fig:star_and_chain_comparison}
\end{figure}

Using these eigenvalues, we evaluate the maximal values of the $n$-locality expressions for both linear chain and star network configurations, assuming equal squeezing for all sources. The resulting violations are shown in Fig.~\ref{fig:star_and_chain_comparison} as functions of the squeezing parameter $K(r)$. In both network topologies, the violation increases monotonically with increasing squeezing and approaches the maximal quantum value $2\sqrt{2}$ in the limit of infinite squeezing ($K \rightarrow 1$).

However, the dependence on the network size is qualitatively different for the
two topologies. For identical TMSV sources, substituting
$\mu_1=1$ and $\mu_2=K(r)$ into the maximal expressions derived in
Sec.~\ref{s3} shows that the linear-chain configuration yields
\begin{equation}
S_{\mathrm{chain}}^{\max}
=
2\sqrt{1 + K(r)^{\,n}} ,
\end{equation}
so that the contribution of the second singular value is raised to a power
that grows with the number of links. Since $0 \le K(r) < 1$ for finite
squeezing, the term $K(r)^{n}$ decreases monotonically with increasing
$n$, leading to a progressive reduction of the attainable violation. This
reflects the multiplicative attenuation of correlations as they propagate
through successive independent links.

In contrast, for the star configuration Eq.~(\ref{eq:star_cv_max}) gives
\begin{equation}
\mathcal{S}_{\mathrm{star}}^{\max}
=
2\sqrt{1 + K(r)^2},
\end{equation}
because
\(
\left(\prod_{i=1}^{n}\mu_1\right)^{2/n}=1
\)
and
\(
\left(\prod_{i=1}^{n}\mu_2\right)^{2/n}=K(r)^2
\),
so the explicit $n$-dependence cancels. Consequently, the maximal violation is
determined solely by the squeezing parameter and remains unchanged as the
number of outer parties increases. This explains why the star-network result yields a single $n$-independent curve in Fig.~\ref{fig:star_and_chain_comparison}, coinciding with the bilocal network scenario. The distinct scaling with network size revealed above directly leads to the
following observations, which summarize the qualitative behavior of the maximal
violations for the two topologies.

\begin{observation}
For linear chain network with identically distributed non-maximally entangled states, the maximal violation of the $n$-locality inequality decreases as the size of the network increases. 
\end{observation}
\begin{observation}
    For star networks with identically distributed states, the maximal $n$-locality violation is independent of the network size.
\end{observation}

\subsection{Resistance to thermal noise}\label{s4B}

The discussions so far have assumed an idealized setting, where the TMSV states are unaffected by any form of environmental disturbance. However, in realistic experimental conditions, the presence of noise is almost unavoidable. We take into account an example of such imperfection by introducing local thermal noise into the system. We explicitly investigate how this noise influences the strength of bilocality violation, showing that it generally diminishes the violation and can, in certain regimes, even eliminate it altogether. Our results obtained below can be readily generalized
to the n-local network scenario, retaining their characteristic features,
since the bilocal network forms the basic building block of both the
linear chain and star network configurations.

Consider the noisy TMSV state,
\begin{align}
    \rho_{*} = p \ket{\zeta}\bra{\zeta} +( 1-p) \left( \sum_{n = 0}^\infty \gamma_n \ket{n}\bra{n} \otimes \sum_{m = 0} \eta_m\ket{m}\bra{m}\right),
\end{align}
where $0\leq p \leq 1$, and $\sum_{n=0}^\infty \gamma_n = \sum_{m=0}^\infty \eta_m = 1$. The local thermal noise parameters are given by
\begin{align}
    \gamma_n &= (1-e^{-\beta_1})e^{-\beta_1 n}, \nonumber \\
    \eta_m &= (1-e^{-\beta_2})e^{-\beta_2 m}
\end{align}
Here, $\beta_i = \frac{1}{k_B T_i}$, representing the inverse temperature of the thermal noise associated with $i$-th mode, $k_B$ is the Boltzmann constant, and the magnitude of the associated energies are assumed to be $1$.

We consider the bilocality scenario, where $\rho_{AB} = \rho_{BC} = \rho_{*}$, with all states having same squeezing value $r$, and all pseudospin observables with $q_i =0$. Using Eq.~(\ref{I}) and Eq.~(\ref{J}), we get,
\begin{align}
    \langle A_x B_0 C_z\rangle_{\rho_{*} \otimes \rho_{*}} &=   \mathcal{N}_1^2 \cos(\theta_1)\cos(\theta_2)
    \\ \langle A_x B_1 C_z\rangle_{\rho_{*} \otimes \rho_{*}} &=  \mathcal{N}_2^2 \sin(\theta_1) \sin(\theta_2)
\end{align}
where 
\begin{align}
    \mathcal{N}_1 &= \left[ p + (1-p)\tanh \frac{\beta_1}{2} \tanh \frac{\beta_2}{2} \right] \nonumber \\
    \mathcal{N}_2 &= pK(r)
\end{align}
with $K(r) = \text{tanh} (2r)$. Using Eq.~(\ref{IJ1}), and Eq.~ (\ref{IJ2}), we can calculate the values of $I$, $J$ and $S$,
\begin{align}
    I_{\rho_{*} \otimes \rho_{*}} &= 4 \cos(\theta_1)\cos(\theta_2) \mathcal{N}_1^2 \nonumber\\
    J_{\rho_{*} \otimes \rho_{*}} &=  4\sin(\theta_1) \sin(\theta_2) \mathcal{N}_2^2 \nonumber\\
    S_{\rho_{*} \otimes \rho_{*}} &= 2 \left[ \mathcal{N}_1\sqrt{\cos(\theta_1)\cos(\theta_2)} +\mathcal{N}_2\sqrt{\sin(\theta_1) \sin(\theta_2)}  \right] 
\end{align} 
The optimal measurement settings, which maximize the value of $S_{\rho_{*} \otimes \rho_{*}}$ are given by,
\begin{align}
\theta_1 = \theta_2 = \theta = \arctan \left( \frac{\mathcal{N}_2}{\mathcal{N}_1} \right),
\end{align}
such that the maximal bilocality quantity $S^{\max}_{\rho_{*}, \rho_{*}}$ reduces to,
\begin{align}
    S^{\max}_{\rho_{*}, \rho_{*}} &= 2\sqrt{\mathcal{N}_1^2+\mathcal{N}_2^2} \nonumber \\
    &=2\sqrt{p^2 K^2 + \left[ p + (1-p) \tanh \frac{\beta_1}{2} \tanh \frac{\beta_2}{2}\right]^2 }
\end{align}

Thus, violation of the bilocality inequality occurs when,
\begin{align}
    p > \frac{-\mathcal{E} +\sqrt{\mathcal{E}^2 - 4\mathcal{DF}}}{2\mathcal{D}}
\end{align}
where,
\begin{align*}
    \mathcal{D} = K^2 +(1-\epsilon)^2, \;
    \mathcal{E} = 2\epsilon(1-\epsilon), \;
    \mathcal{F} = \epsilon^2-1, 
\end{align*}
with $\epsilon = \tanh \frac{\beta_1}{2} \tanh \frac{\beta_2}{2}.$

\begin{figure}
    \centering
    \includegraphics[width=8.7cm]{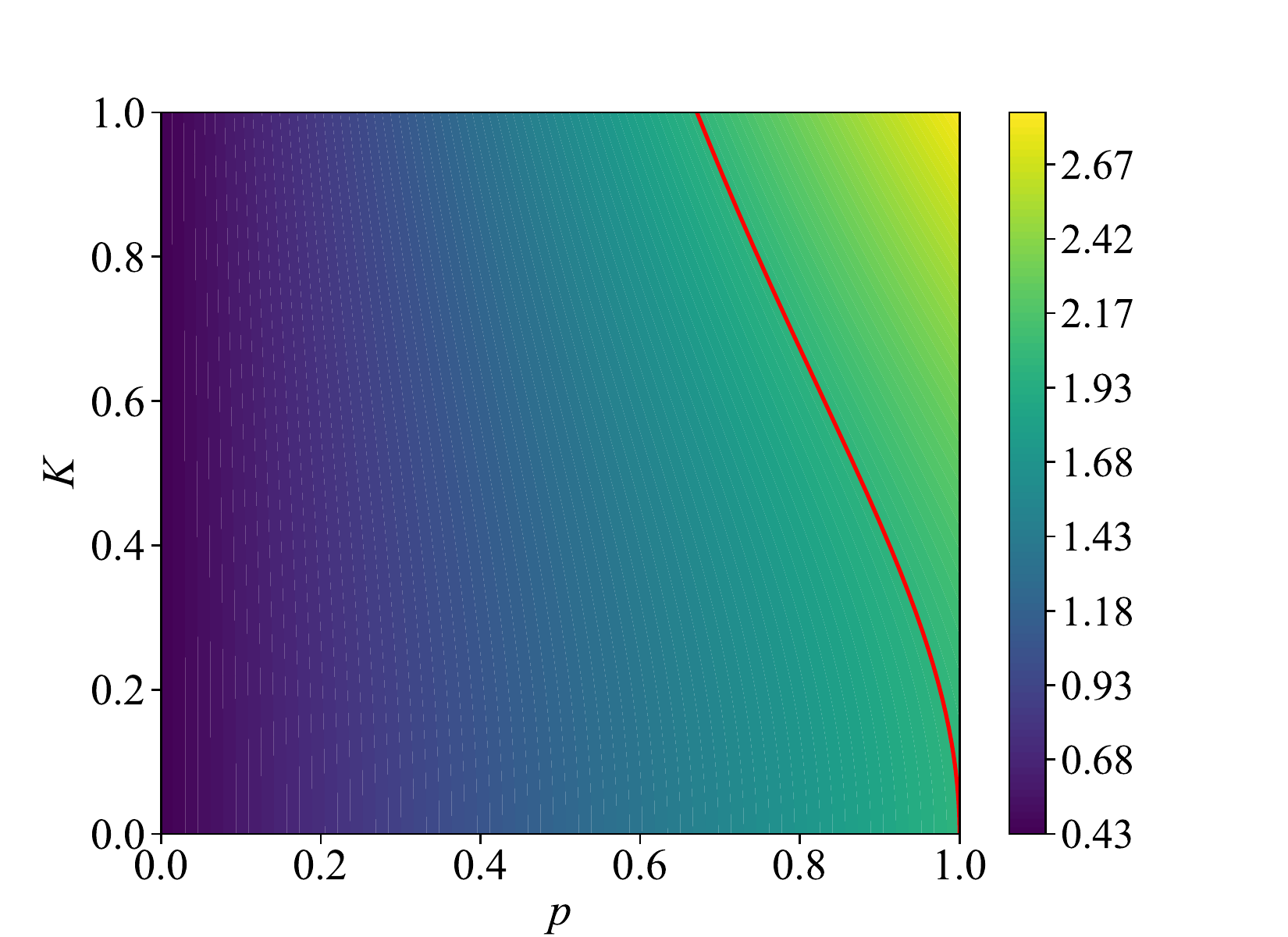}
    \caption{
        \justifying
         $S^{\max}_{\rho_{*}, \rho_{*}}$ is plotted as a function of $p$ and $K$, for fixed inverse temperature parameters $\beta_1 = \beta_2 = 1$. The red contour line corresponds to $ S^{\max}_{\rho_{*}, \rho_{*}} = 2$, separating the bilocal ($S^{\max}_{\rho_{*}, \rho_{*}} \leq 2$) and nonbilocal ($S^{\max}_{\rho_{*}, \rho_{*}} > 2$) regions in the parameter space.
         }
    \label{fig:2}
\end{figure}

\begin{figure}
    \centering
    \includegraphics[width=8.7cm]{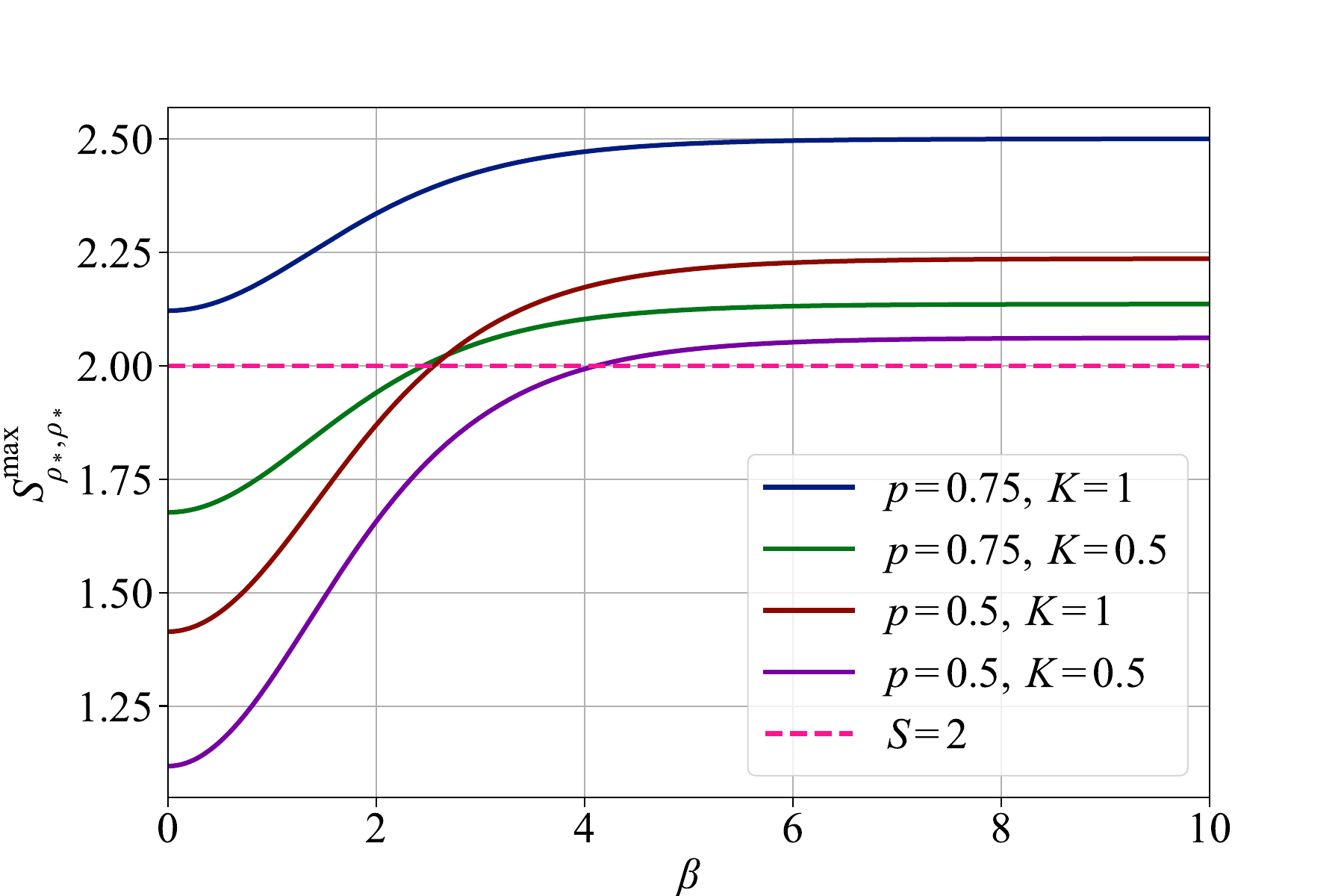}
    \caption{
        \justifying
        Variation of the maximized bilocality expression $S^{\max}_{\rho_{*}, \rho_{*}}$ for the noisy TMSV state as a function of the inverse temperature parameter $\beta$, for different values of $(p, K)$. The plot illustrates that at low $\beta$ (i.e., high temperature), nonbilocal correlations are suppressed, and disappear below a certain threshold. As $\beta$ increases (low-temperature regime), the impact of thermal noise diminishes, and the nonbilocality becomes more robust.
    }
    \label{fig:3}
\end{figure}

To visualize the effect of local thermal noise on bilocality violation, we examine the maximal bilocality expression $S^{\max}_{\rho_{*}, \rho_{*}}$ across relevant state and noise
parameters. In Fig.~\ref{fig:2}, we fix the thermal environment to $\beta_1=\beta_2=1$ and study the dependence of $S^{\max}_{\rho_{*}, \rho_{*}}$ on the noise probability $p$ and the
entanglement parameter $K=\tanh(2r)$. The plot identifies the critical boundary separating bilocal and nonbilocal regions, illustrating how squeezing and noise jointly determine the network behavior. In
Fig.~\ref{fig:3}, we instead vary the inverse temperature $\beta_1=\beta_2=\beta$, revealing how thermal fluctuations influence the bilocality expression. As $\beta$ decreases (higher temperature),
the violation diminishes and eventually disappears for certain states, demonstrating the fragility of nonclassical correlations under thermal noise. Conversely, stronger entanglement and lower noise fractions allow the system to retain nonbilocality over a wider temperature range. These features motivate the following observations summarizing the thermal suppression and robustness of bilocality violations.

\begin{observation}
As $\beta$ decreases (i.e.\ as the temperature increases), nonbilocality is suppressed (see Fig.~\ref{fig:3}).
\end{observation}

\begin{observation}\label{temp}
Bilocality violation is guaranteed independently of the thermal
parameters whenever
\begin{equation}
K>\sqrt{\frac{1}{p^2}-1}.
\end{equation}
Under this condition, the bilocality expression remains violated
throughout the entire physically accessible range of noise temperatures $(\beta_1, \beta_2)$,
ensuring persistence of nonbilocal correlations even in the presence
of arbitrarily strong thermal noise (see Appendix~\ref{s9_obs4} for the derivation).
\end{observation}

\section{Network nonlocality with non-Gaussian states} \label{s5}

Despite their broad applicability, Gaussian states have a key limitation: measurement statistics involving only Gaussian operations can always be efficiently simulated classically \cite{Bartlett2002}. This underscores the importance of non-Gaussian states, which are crucial for achieving universal CV quantum computation \cite{Lloyd1999}. Here we explore nonbilocality using three prominent classes non-Gaussian states: single photon subtracted two-mode squeezed vacuum state, entangled coherent state and CV Werner state.


\subsection{Enhanced bilocality violation for  photon subtracted squeezed vacuum states}

Violation of bilocality inequality with TMSV has been studied in Sec.~\ref{s4}. TMSV can be de-Gaussified through local photon addition or subtraction on individual modes. From an experimental perspective, photon subtraction is generally more feasible than photon addition, as the latter typically requires an auxiliary photon pumping apparatus~\cite{Kim2008}. Although both operations can yield enhanced nonlocality in theory, photon subtraction is often the method of choice in practical implementations. In what follows, we investigate how single photon subtraction influences the extent of violation of bilocality inequality. 

We consider three types of photon subtracted states, {\it viz.}, (a) asymmetric photon subtraction (single photon from one mode), (b) symmetric photon subtraction (single photon from both modes), and (c) coherent superposition of photon subtraction across modes. In each case, we analyze two network configurations, {\it viz.}, (i) hybrid scenario where Alice-Bob share the photon subtracted state, while Bob-Charlie share a TMSV state, and
(ii) symmetric scenario where both shared states are photon subtracted and identical. We assume the squeezing parameter \(r\) to be the same for all involved states. The corresponding maximization procedures follow the prescription detailed in Sec.~\ref{s3}.


\subsubsection{Asymmetric Photon Subtraction: Single Mode Operation} \label{s5A}

We first study the impact of asymmetric photon subtraction on the violation of the bilocality inequality. Specifically, we consider the case where a single photon is subtracted from one mode of a TMSV state.

Applying the annihilation operator \((\mathbb{I} \otimes \hat{b})\) on the second mode of a TMSV yields the normalized state,
\begin{align}
    \ket{\xi_1} &= \left(\frac{1-\lambda^2}{\lambda}\right)\sum_{n=1}^{\infty} \lambda^n \sqrt{n} \ket{n, n-1} \nonumber\\
    & = (1-\lambda^2) \sum_{n=0}^{\infty} \lambda^n \sqrt{n+1} \ket{n+1, n}
\end{align}
where \(\lambda = \tanh r\) and \(\hat{b}\) is the annihilation operator for the second mode. 
\vspace{2mm}

\textbf{\textit{Configuration A1:}} In the first scenario, Alice and Bob share the photon subtracted state \(\ket{\xi_1}\), while Bob and Charlie share the standard TMSV state \(\ket{\zeta}\). The maximal value of the bilocality expression is given by
\begin{align}
S^{\max}_{\xi_1,\zeta} = 2 \sqrt{\nu_1^{(a)} \mu_1 + \nu_2^{(a)} \mu_2}.
\end{align}
Here $\{\mu_i\}$ and $\{\nu^{(a)}_i\}$ are two largest eigenvalues of the corresponding correlation matrices for the states $\ket{\zeta}$ and $\ket{\xi_1}$ respectively. For further details including expressions of all parameters, see Appendix~\ref{s8A}.
\vspace{2mm}

\textbf{\textit{Configuration A2:}} In the second configuration, both shared states are identical and given by the photon subtracted state \(\ket{\xi_1}\).  
In this case, the maximal value of the bilocality expression is given by,
\begin{align}
S^{\max}_{\xi_1,\xi_1} = 2 \sqrt{ (\nu_1^{(a)})^2 + (\nu_2^{(a)})^2 }.
\end{align}
Our  results for bilocality violation versus the squeezing parameter for both the above cases are presented in Fig.~\ref{fig:spssv1}. 
\begin{observation}
    The configuration A2 leads to more bilocality violation compared to the configuration A1 for any finite squeezing.
\end{observation}

\vspace{2mm}
\noindent

\subsubsection{Symmetric Photon Subtraction: Dual-Mode Operation} \label{s5B}

We next investigate the impact of symmetric photon subtraction on the maximal violation of the bilocality inequality. Specifically, we investigate the case where a single photon is subtracted from both modes of a TMSV.

Applying the annihilation operators \((\hat{a} \otimes \hat{b})\) on the TMSV yields the normalized two-mode state
\begin{align}
    \ket{\xi_2} &= \left( \frac{1-\lambda^2}{\lambda \sqrt{1+\lambda^2}} \right) \sqrt{1 - \lambda^2} \sum_{n=1}^{\infty} \lambda^n n \ket{n-1, n-1} \nonumber \\
                &= (1 - \lambda^2) \sqrt{\frac{1-\lambda^2}{1+\lambda^2}} \sum_{n=0}^{\infty} \lambda^n (n+1) \ket{n, n},
\end{align}
where \(\lambda = \tanh r\), and \(\hat{a}\), \(\hat{b}\) denote annihilation operators for the first and second mode, respectively.

\vspace{2mm}
\textbf{\textit{Configuration B1:}} In this configuration, the entangled state shared between Alice and Bob is \(\ket{\xi_2}\), while Bob and Charlie share the original TMSV state \(\ket{\zeta}\).
The resulting maximal bilocality expression is given by,
\begin{align}
S^{\max}_{\xi_2,\zeta} = 2 \sqrt{\nu_1^{(s)} \mu_1 + \nu_2^{(s)} \mu_2}.
\end{align}

\vspace{2mm}
\textbf{\textit{Configuration B2:}} Here, both shared states are taken to be \(\ket{\xi_2}\), making this a fully symmetric setting. The maximum bilocality expression in this configuration is 
\begin{align}
S^{\max}_{\xi_2,\xi_2} = 2 \sqrt{ (\nu_1^{(s)})^2 + (\nu_2^{(s)})^2 }.
\end{align}
Our  results for bilocality violation versus the squeezing
parameter for both the above cases are presented in Fig.~\ref{fig:spssv1}. For a more detailed discussion, see Appendix~\ref{s8A}. 

\begin{observation}\label{B2vsA2}
Symmetric photon subtraction (B2) produces a stronger violation of the
bilocality inequality than the asymmetric scheme (A2) for
$0 < \lambda < 0.616$. Beyond this crossover point, i.e.,
for larger squeezing parameters, the ordering reverses and the
asymmetric scheme yields the stronger violation. An analogous behaviour
is observed when comparing configurations B1 and A1 (see
Appendix~\ref{s9_obs6} for the derivation).
\end{observation}
{\it Explanation.}
This crossover can be understood by examining how photon subtraction reshapes the photon-number distribution of the TMSV state in a way that affects pseudospin correlations. At low squeezing, the TMSV state is only weakly entangled and dominated by low photon-number components, so symmetric subtraction enhances correlated higher-photon sectors in both modes and generates stronger non-Gaussian correlations, leading to the ordering $B2 > A2$ and $B1>A1$. As the squeezing increases, however, the TMSV already possesses strong multi-photon correlations, and the nonlocality test becomes sensitive not simply to entanglement strength, but to how different number sectors contribute to the pseudospin observables. In this regime symmetric subtraction redistributes weight too strongly toward very high photon numbers, which are not optimally aligned with these correlators, whereas single-mode subtraction introduces a milder imbalance that preserves useful correlations while still inducing non-Gaussianity, resulting in a reverse of the order i.e. $A2>B2$ and $A1>B1$ beyond the crossover. Importantly, both photon-subtracted states continue to outperform the Gaussian TMSV even at large squeezing, reflecting the well-known advantage of non-Gaussian features for pseudospin-based nonlocality tests, where higher-order coherences produced by subtraction enable stronger violations than those achievable with Gaussian states alone.

\begin{observation}
     The asymmetric subtraction applied to both states (A2) still outperforms the symmetric subtraction applied to only one state (B1), highlighting the benefit of distributing non-Gaussianity across the network (see Fig.~\ref{fig:spssv1}).
\end{observation}

\subsubsection{Coherent Superposition of Photon Subtractions across Modes} \label{s5C}

Here we consider an explicit example of a non-Gaussian state that is obtained with coherent superposition of two asymmetric single photon subtracted states. This photon subtraction operation can be written as, 
\begin{equation}
    \mathcal{G} = a \otimes I + (-1)^k I \otimes b
\end{equation}
where $k\in \{0,1\}$. It is also assumed that the specific mode from which the photon is subtracted remains unknown. The state is represented as \cite{chowdhury2014, mallick2025}, 
\begin{align}
\ket{\xi_{3}}
&= \left( \frac{1-\lambda^2}{\sqrt{2}} \right)
\sum_{n=1}^{\infty}\lambda^{\,n-1}\sqrt{n}
\Big[
\ket{n-1,n}
+(-1)^k\ket{n,n-1}
\Big]
\nonumber\\
&= \left( \frac{1-\lambda^2}{\sqrt{2}} \right)
\sum_{n=0}^{\infty}\lambda^{\,n}\sqrt{n+1}
\Big[
\ket{n,n+1}
+(-1)^k\ket{n+1,n}
\Big],
\end{align}
with $\lambda = \tanh{r}, \hspace{0.1cm} r >0$. Here we consider the state with $k=0$.

It is known that CV entangled states of the form 
\(\ket{\psi} = \sum_{n=0}^{\infty} \mathcal{C}_n \ket{n, n+\bar{n}}\) or 
\(\ket{\psi} = \sum_{n=0}^{\infty} \mathcal{C}_n \ket{n+\bar{n}, n}\) 
do not exhibit a violation of the Bell-CHSH inequality when \(\bar{n}\) is an odd integer and the pseudospin measurement parameters \(q_i\) are chosen to be zero \cite{zhang2011}. However, we demonstrate that this limitation does not hold when coherent superpositions of such states are considered.

\vspace{2mm}
\textbf{\textit{Configuration C1:}} In this configuration, the entangled state shared between Alice and Bob is \(\ket{\xi_3}\), while Bob and Charlie share the original TMSV state \(\ket{\zeta}\).
The resulting maximal bilocality expression is given by (see Appendix~\ref{s8A}),
\begin{align}
S^{\max}_{\xi_3,\zeta} = 2 \sqrt{\nu_1^{(s)} \mu_1 + \nu_2^{(s)} \mu_2}.
\end{align}

\begin{figure*}[t]
    \centering
    \begin{subfigure}{0.48\textwidth}
        \centering
        \includegraphics[width=\linewidth]{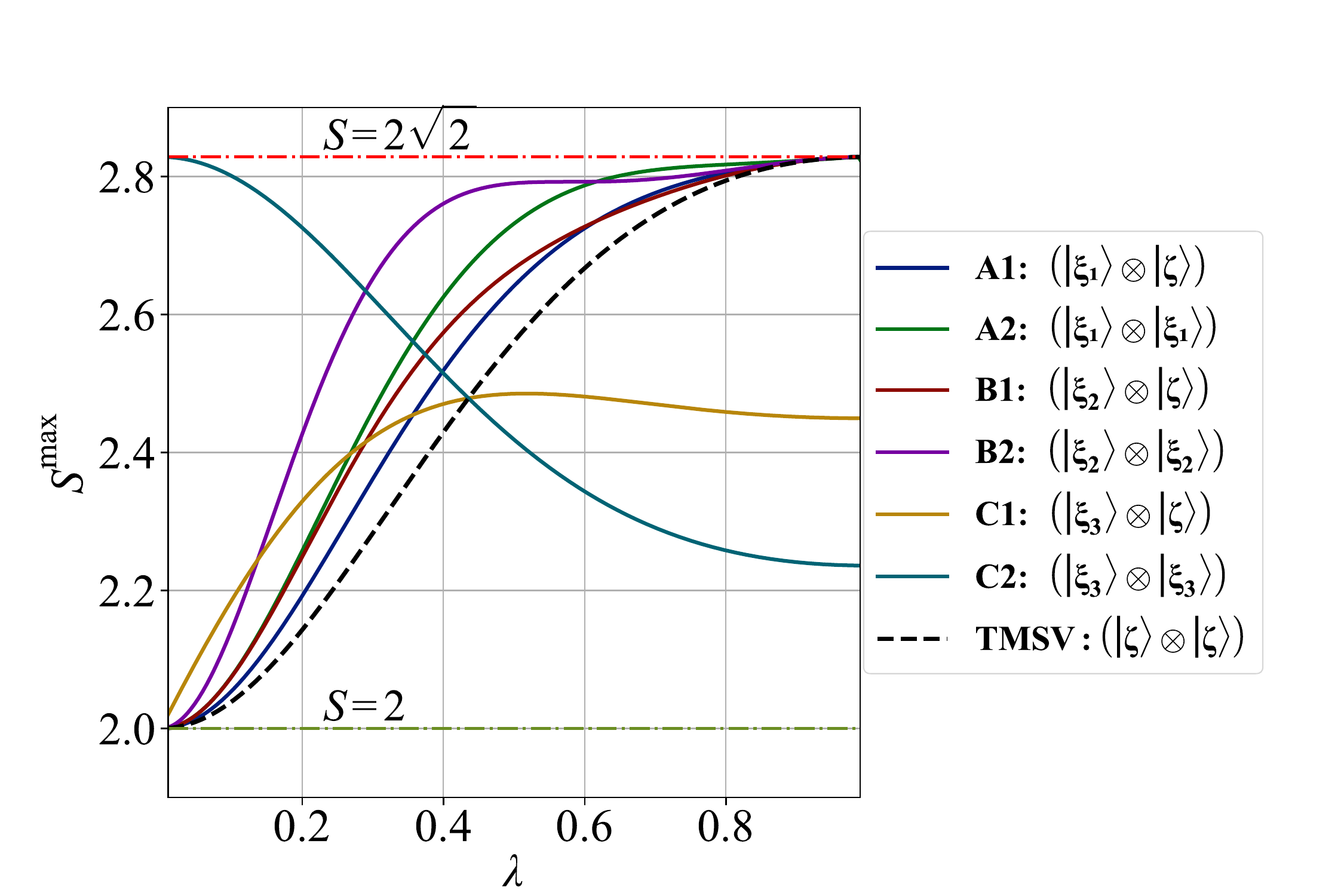}
        \caption{}
        \label{fig:spssv1}
    \end{subfigure}
    \hfill
    \begin{subfigure}{0.48\textwidth}
        \centering
        \includegraphics[width=\linewidth]{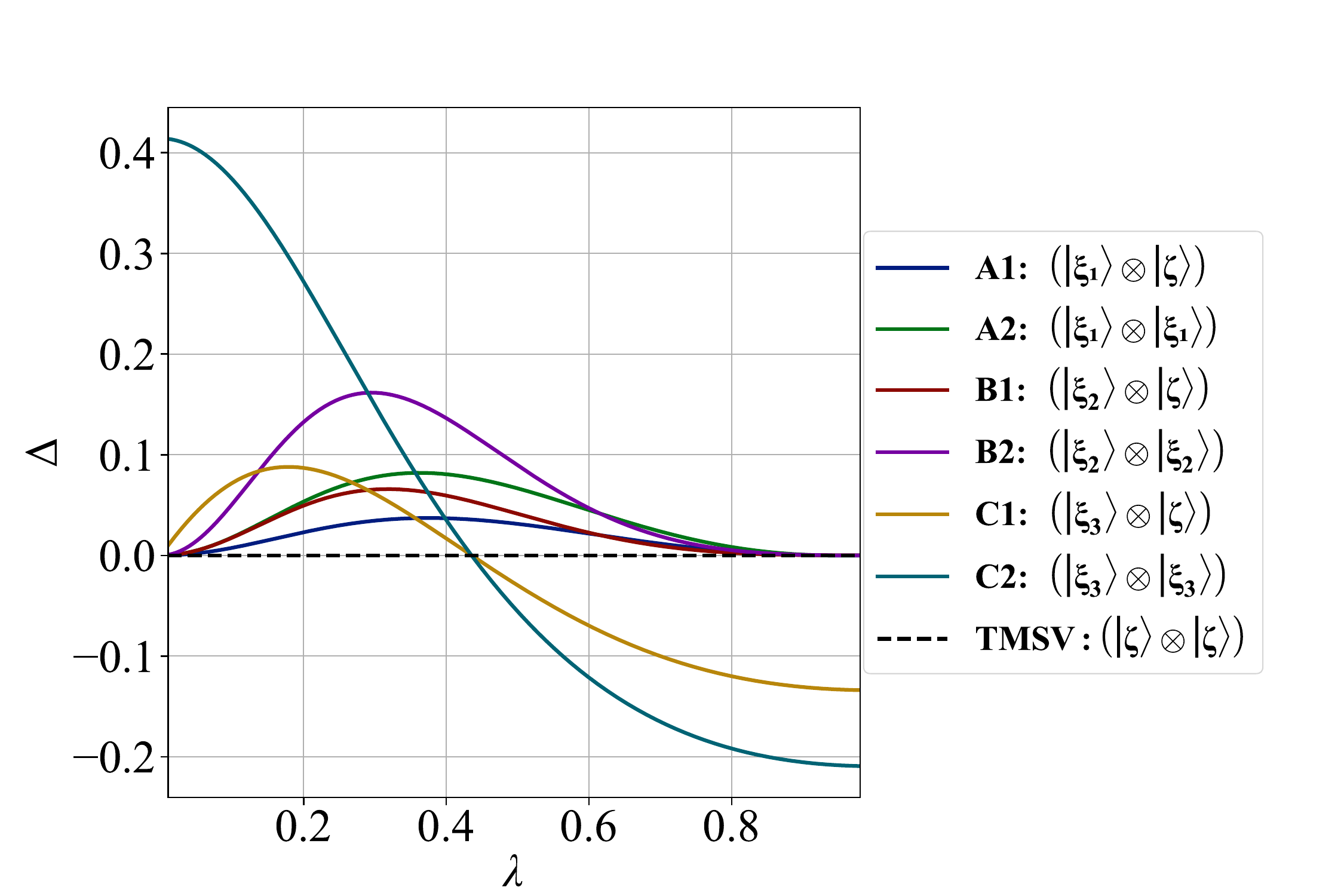}
        \caption{}
        \label{fig:spssv2}
    \end{subfigure}
    
\caption{
\justifying
Maximal bilocality violation for different configurations involving single-photon–subtracted squeezed vacuum (SPSSV) states as a function of the squeezing parameter $\lambda$, with the TMSV case taken as a reference. 
(a) Maximized bilocality expression $S^{\max}$, showing that configurations A1, A2, B1, and B2 exhibit enhanced violations over the full range of $\lambda$, while configurations C1 and C2 show enhancement only up to $\lambda=0.4354$. In configuration C2, where both entangled pairs are in the state $\ket{\xi_3}$, the maximal quantum value $2\sqrt{2}$ is attained at $\lambda=0$. 
(b) Relative enhancement $\Delta$ of the bilocality violation with respect to the TMSV reference scenario.
}
    \label{fig:spssv_combined}
\end{figure*}

\vspace{2mm}
\textbf{\textit{Configuration C2:}} Here, both shared states are taken to be \(\ket{\xi_2}\), making this a fully symmetric setting. 
The maximum bilocality expression in this configuration is given by

\begin{align}
S^{\max}_{\xi_3,\xi_3} = 2 \sqrt{ (\nu_1^{(c)})^2 + (\nu_2^{(c)})^2 }.
\end{align}
It may be noted from Fig.~\ref{fig:spssv1} that all photon subtracted cases yield higher bilocality violation compared to the TMSV  in the low squeezing regime. Moreover, the case with coherent superposition of subtraction of photons (C2) performs better in bilocality violation compared to the other two cases of symmetric and asymmetric photon subtraction for low squeezing (see Fig.~\ref{fig:spssv1}). 

\begin{observation}\label{c1_enhancement}
Both configurations C1 and C2 exhibit stronger bilocality violations than the TMSV state within the squeezing range $0<\lambda<0.4354$, with configuration C2 providing the largest enhancement. Beyond this threshold, neither configuration yields an advantage over the TMSV, and configuration C2 performs worse than C1 (see Appendix~\ref{s9_obs8} for details).
\end{observation}

{\it Explanation.}
This observation provides a concrete example that non-Gaussian operations do not universally enhance nonlocal correlations. In the low-squeezing regime the TMSV state itself carries only weak nonlocal correlations, and the observed enhancement arises primarily from the action of the non-Gaussian operation $\mathcal{G}$. The coherent subtraction reshapes the photon-number amplitudes and introduces interference between neighboring sectors, which are strongly aligned with the structure probed by pseudospin measurements, thereby increasing the violation. As the squeezing increases, however, the underlying Gaussian correlations of the TMSV become dominant and already support substantial violations. In this regime the additional reshaping induced by $\mathcal{G}$ no longer matches the measurement structure optimally and instead redistributes weight into sectors that contribute less effectively to the correlators. Consequently, the advantage diminishes and eventually disappears, illustrating that the usefulness of non-Gaussian operations for nonlocality tests is regime dependent rather than universal.

\begin{observation}
For the configuration C2 with the state \(\ket{\xi_3}\), the bilocality expression attains the maximum, 
\(S^{\max}_{\xi_3,\xi_3} = 2\sqrt{2}\), in the limit \(\lambda \rightarrow 0\), corresponding to zero squeezing. This indicates a maximal violation of the bilocality inequality, (as well as CHSH inequality with $\ket{\xi_3}$) despite the absence of squeezing  (see Fig.~\ref{fig:spssv1}).
\end{observation}
{\it Explanation.}  It is important to emphasize that \(\ket{\xi_3}\) is not a maximally entangled state in the full infinite-dimensional Hilbert space when \(\lambda \rightarrow 0\), as the von Neumann entropy remains finite when traced over one mode. This apparent paradox is resolved by noting that, in the limit $\lambda \rightarrow 0$, the state $\ket{\xi_3}$ effectively resides in a two-dimensional subspace. Within this subspace, it is indeed a maximally entangled state, as it takes the form:
\begin{align}
    \ket{\xi_3(\lambda \rightarrow 0)} = \frac{1}{\sqrt{2}}(\ket{01} + \ket{10}),
\end{align}
which belongs to the parity based Bell states defined in Eq.~\eqref{bell_states}, with ${\cal A}_0 = {\cal B}_{0}=1$ and ${\cal A}_n={\cal B}_{n}=0$ for any $n \neq 0$. Since pseudospin measurements probe precisely this parity structure, the bilocality expression reaches its algebraic maximum even in the absence of squeezing. Thus the maximal violation originates not from large CV entanglement, but from the effective qubit embedding induced by the coherent superpositions of photon subtraction at zero squeezing.

The enhancement in the maximal violation of the bilocality inequality due to photon subtraction can be quantified using the measure
\begin{align}
    {\Delta} = \frac{S^{\max}_{\psi_{AB},\phi_{BC}} - S^{\max}_{\zeta,\zeta}}{S^{\max}_{\zeta,\zeta}}.
\end{align}

In this expression, \(S^{\max}_{\psi_{AB},\phi_{BC}}\) denotes the maximal bilocality violation for a given configuration of non-Gaussian entangled states \(\ket{\psi_{AB}} \otimes \ket{\phi_{BC}}\), while \(S^{\max}_{\zeta,\zeta}\) corresponds to the maximal violation for the reference configuration comprising TMSV states, \(\ket{\zeta} \otimes \ket{\zeta}\). This measure allows us to systematically compare the advantage gained through non-Gaussian operations (see Fig.~\ref{fig:spssv2}).

\subsection{Entangled coherent states (ECS)}

The entangled coherent state (ECS) \cite{Sanders_1992_ECS} is an important non-Gaussian CV entangled state. Such states, also labeled as "cat" states are regarded to be of foundational as well as practical relevance \cite{nandi2024, Hoshi2025}. Several applications in quantum information processing have been studied utilizing ECS \cite{Van_Enk_ECS, DeFabritiis2023_ECS}. Here we consider the asymmetric ECS ($|{\rm ECS}\rangle$), which can be defined as\cite{dichotomic_nonlocality},
\begin{eqnarray}
|{\rm ECS}\rangle_\alpha ={\cal N}(|\alpha\rangle|-\alpha\rangle
 -|-\alpha\rangle|\alpha\rangle), \label{ecs1} \;\;\;
\end{eqnarray}
where, $|\alpha\rangle =
e^{-|\alpha|^2/2}\sum_{n=0}^{\infty}\frac{\alpha^n}{\sqrt{n!}}|n\rangle$, $|\alpha\rangle$ is a coherent state with $\alpha\neq0$, we have considered $\alpha$ to be real for simplicity. ${\cal N}$ is a normalization factor given by,
\begin{equation}
   { \cal {N}} = \frac{1}{\sqrt{2(1-e^{-4\alpha^2})}}
\end{equation}
The expression for the maximum bilocality $S^{\max}$ is given by
\begin{equation}
    S^{\max}_{\alpha, \beta} = 2\sqrt{1+Q^2}
\end{equation}
where $Q$ is a function of the coherent state parameters $\alpha,\beta$
(see, Appendix~\ref{s8B} for the derivation).

\begin{figure}
    \centering
    \includegraphics[width=8.7cm]{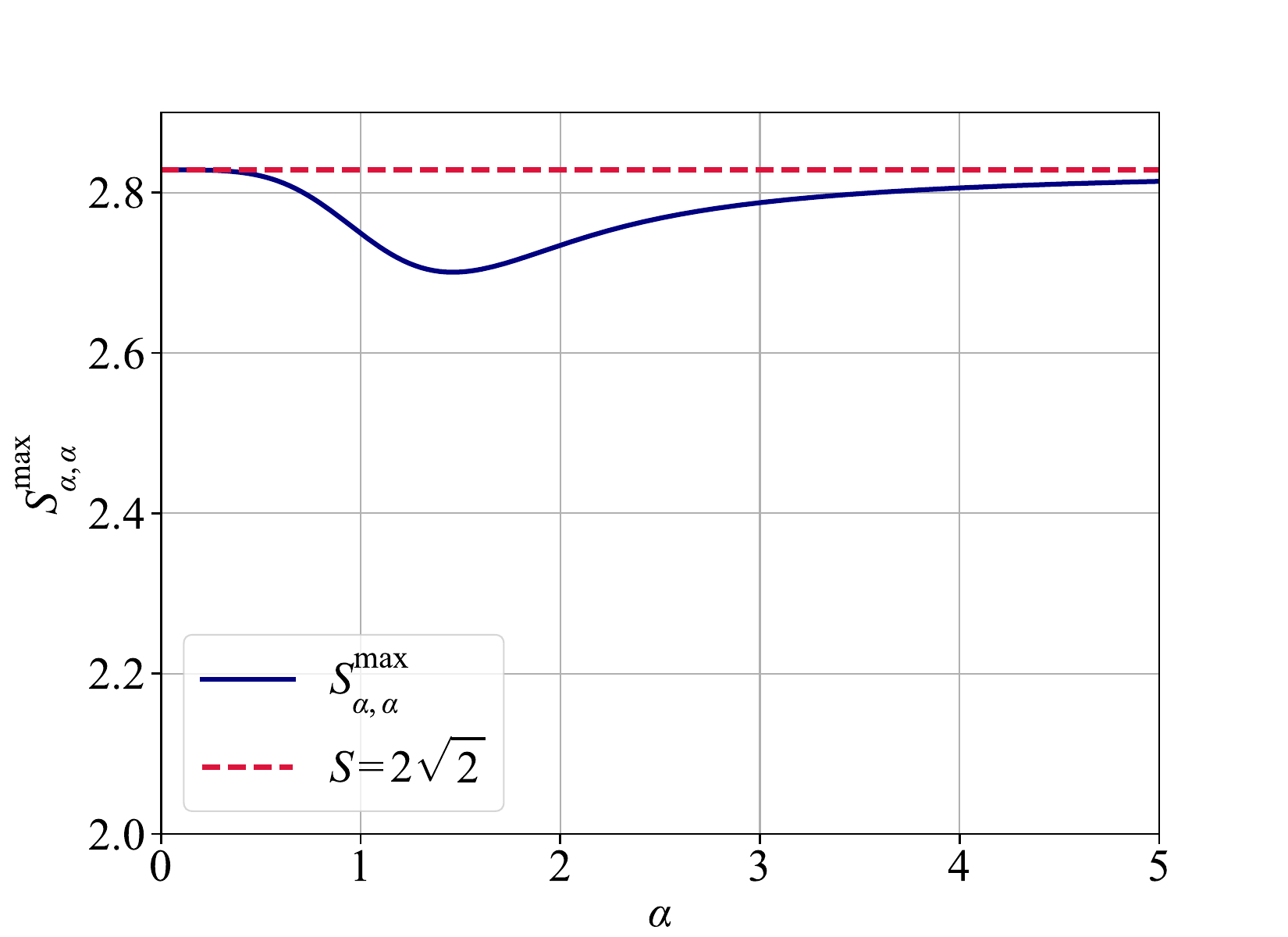}
    \caption{
    \justifying
        Variation of the maximized bilocality expression $S^{\max}_{\alpha, \alpha}$ for ECS as a function of $\alpha$ (blue curve).  $S^{\max}_{\alpha, \alpha}$ reaches maximum $(=2\sqrt{2})$ when $\alpha \rightarrow 0$ or $\alpha \rightarrow \infty$.
    }
    \label{fig:ecs}
\end{figure}

\begin{observation}
ECS always violates the bilocality inequality Eq.~(\ref{bilocal}) as, $0<Q(\alpha, \beta)<1$, and the degree of violation is completely specified by magnitude of $Q(\alpha, \beta)$. $Q(\alpha, \beta)$ approaches $1$  when $\alpha, \beta \rightarrow 0$ (but $ \alpha, \beta \neq 0 $) giving maximal violation of the bilocality inequality, and when $\alpha, \beta \rightarrow \infty $ (see Fig.~\ref{fig:ecs}). 
\end{observation}

\subsection{CV Werner state}

In the DV regime, the Werner state \cite{werner_1989} is a convex mixture of a maximally entangled state and the maximally mixed state:
\begin{equation}
    \rho_{\text{Werner}} = p \ket{\Psi_d}\bra{\Psi_d} + \frac{1-p}{d^2} I_1 \otimes I_2,\;\; 0\leq p\leq1,
\end{equation}
where $\ket{\Psi_d} = \frac{1}{\sqrt{d}} \sum_{i=1}^d \ket{i}_1\ket{i}_1$. Extending this concept to CV systems is nontrivial due to the absence of a proper maximally mixed state in infinite dimensions. An operational analogue can be defined by mimicking the DV structure. Since the reduced states of a TMSV are thermal states, a natural CV analogue involves mixing the TMSV with a product of two identical thermal states \cite{CV_Werner},
\begin{equation}\label{rhot}
\rho_{T} = (1-\lambda_2^2)^2 \sum_{m,n=0}^\infty \lambda_2^{2(m+n)} \ket{m}\bra{m} \otimes \ket{n}\bra{n},
\end{equation}
where $\lambda_2 = \tanh s$ and $\langle n\rangle_T = \sinh^2 s$.

Thus, the CV Werner state takes the form
\begin{equation}\label{rhow}
\rho_W = p\,\rho_{\text{TMSV}} + (1-p)\,\rho_T,\quad 0 \leq p \leq 1,
\end{equation}
where $\rho_{\text{TMSV}} = \ket{\zeta}\bra{\zeta}$, and $\ket{\zeta}$ is the TMSV state (\ref{nopa}). This defines a class of states parametrized by $(p, r, s)$. For simplicity, we take $r = s$, reducing to two parameters $(p, r)$. In the limit $r \to \infty$, the CV Werner state approaches a mixture of a maximally entangled EPR state and a maximally mixed state, closely resembling the DV case.

\begin{figure*}[t]
    \centering
    \begin{subfigure}{0.48\textwidth}
        \centering
        \includegraphics[width=\textwidth]{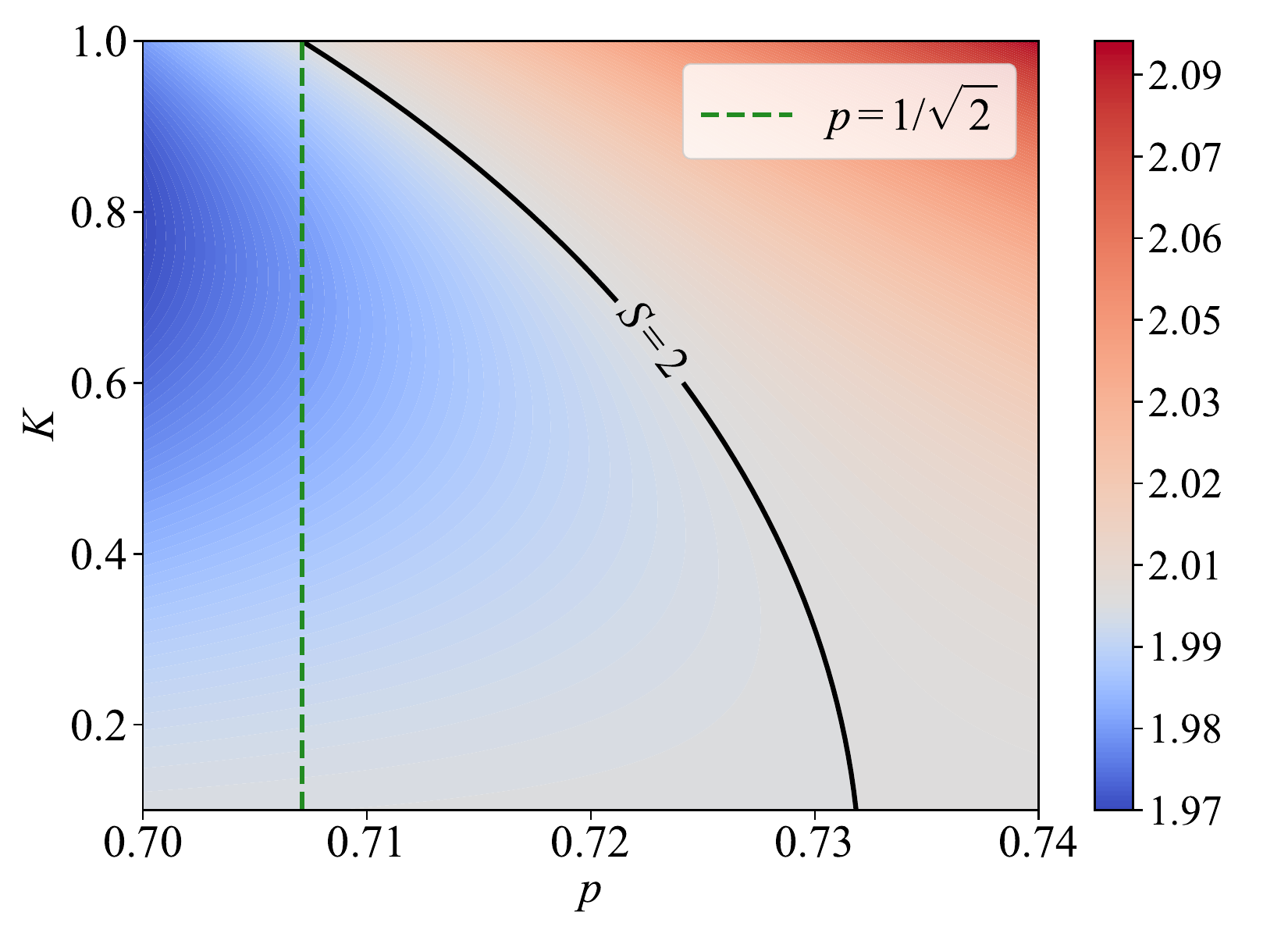}
        \caption{}
        \label{fig:werner_a}
    \end{subfigure}
    \hfill
    \begin{subfigure}{0.48\textwidth}
        \centering
        \includegraphics[width=\textwidth]{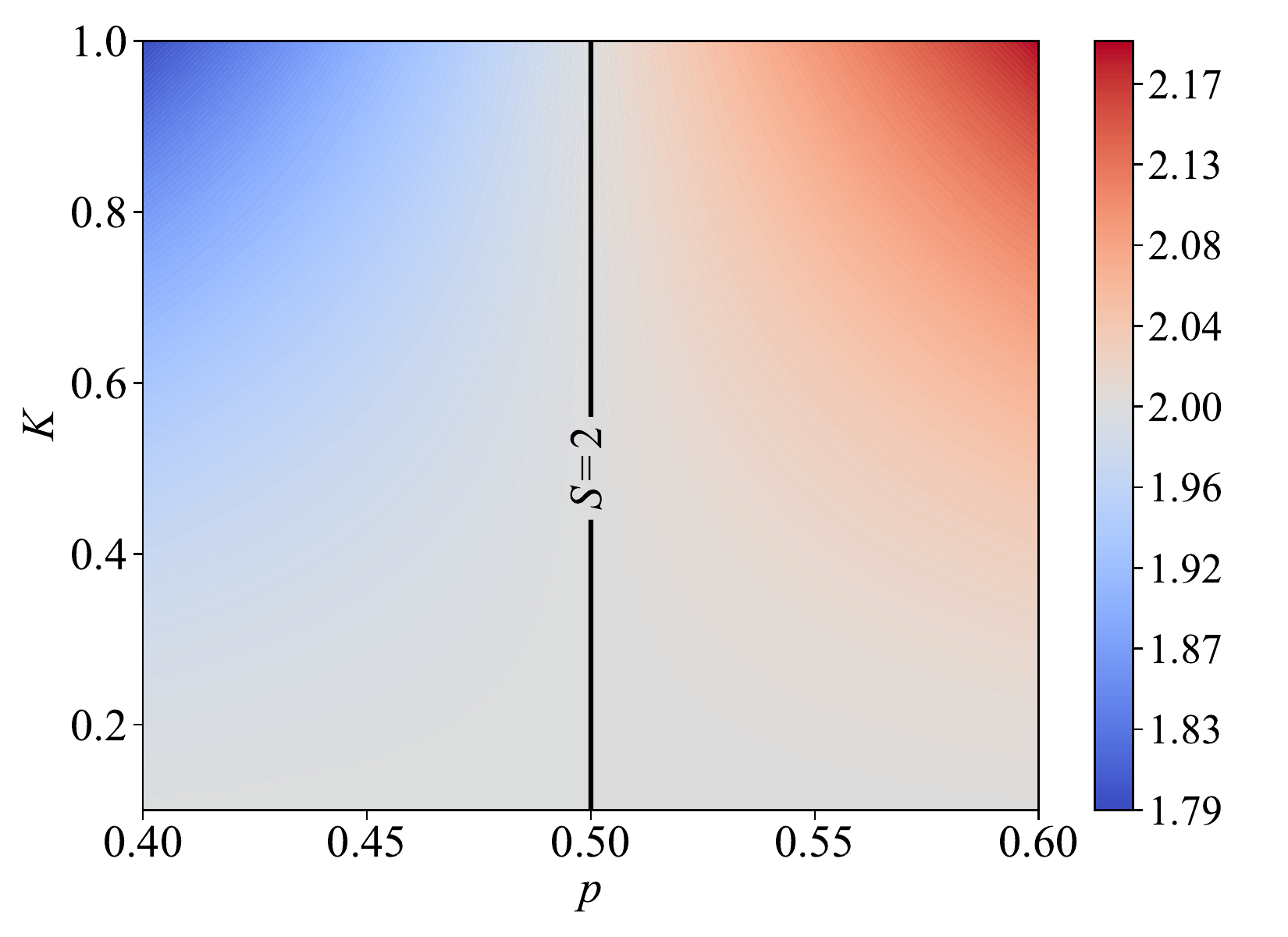}
        \caption{}
        \label{fig:werner_b}
    \end{subfigure}

    \caption{
    \justifying
    Maximal bilocality value for a CV Werner state as a function of the noise parameter \( p \) and squeezing parameter \( K \).
    (a) Symmetric configuration with \( p_1 = p_2 = p \).
    (b) Asymmetric configuration with \( p_1 = 1 \) and \( p_2 = p \).
    In both cases, the black curve corresponds to \( S^{\max} = 2 \), separating the nonbilocal region from the bilocal one.
    For \( K = 1 \), violation in (a) requires \( p > 1/\sqrt{2} \), while in (b) it occurs for any non-zero \( K \) when \( p > 1/2 \).
    }
    \label{fig:werner_combined}
\end{figure*}

We analyze the bilocality violation in the following two scenarios:

\textbf{\textit{Case 1:}} When both Werner-type resource states have equal noise parameters, i.e., $p_1 = p_2 = p$. We obtain (see, Appendix~\ref{s8C} for further details),
\begin{align}
     S^{\max}_{\rho_{W_1}, \; \rho_{W_2}} &(p_1 = p_2 = p)
     = 2 \times \{ p^2 (1+K^2) \nonumber\\
     &+2p (1-p)(1-K^2) + (1-p)^2 (1-K^2)^2 \}^\frac{1}{2}
\end{align}
Violation of the bilocality inequality occurs when,
\begin{align}
    &p > \frac{-E + \sqrt{E^2 - 4D(F - 1)}}{2D} \notag     
\end{align}
where, $D = 3K^2 - 1 + (1 - K^2)^2,\quad 
    E = 2K^2(1 - K^2),\quad F = (1 - K^2)^2 - 1$ .
    
\begin{observation}
    For the extremal case \( K = 1 \), with $p_1 = p_2 = p$, CV Werner state shows nonbilocality when \( p > \frac{1}{\sqrt{2}} \), matching the known threshold for nonbilocality with two-qubit Werner states (see Fig.~\ref{fig:werner_a}).
\end{observation}

\textbf{\textit{Case 2:}} Now we set, \( p_1 = 1 \) and \( p_2 = p \), we find
\begin{align}
    S^{\max}_{\rho_{W_1}, \; \rho_{W_2}}(p_1 = 1,\; p_2 = p) = 2 \sqrt{1 + (2p-1)K^2}.
\end{align}
\begin{observation}
    With $p_1=1, \;p_2=p$, CV Werner states violate bilocality inequality whenever $p>1/2$, for any nonzero squeezing (i.e. $K>0$) (see Fig.~\ref{fig:werner_b}).
\end{observation}


\section{Schematics for experimental realization}\label{s6}

In this section, we outline an experimentally feasible approach to realizing nonbilocal correlations using spatial parity observables. Building on the spatial-parity–based Bell test demonstrated in Ref.~\cite{parity_expt}, we extend this framework to a bilocal network scenario. Moreover, the same strategy can be straightforwardly generalized to larger network topologies, including linear-chain and star configurations.

\begin{figure*}[t]
\includegraphics[width=13cm]{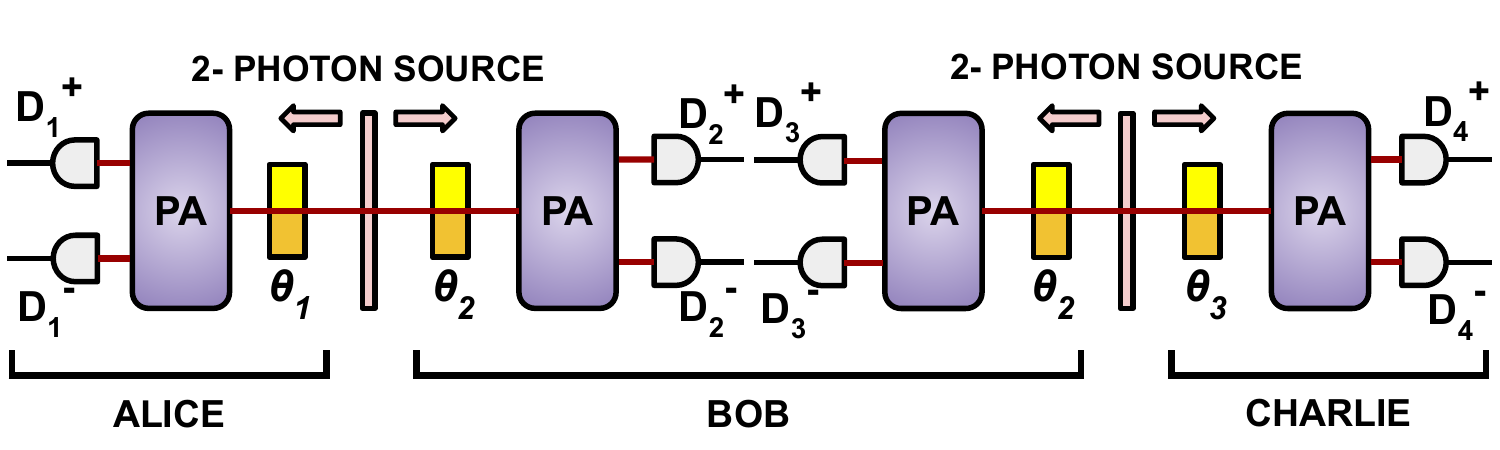}
\caption{Schematic configuration for testing nonbilocal correlations using spatial parity measurements. PA denotes the parity analyzer, $\theta$ the parity rotator, and D the detector.}
\label{fig:bilocality_expt}
\end{figure*}

Pseudospin observables, together with the associated rotation and projection operators, admit a natural physical realization within the spatial parity framework~\cite{spatial_parity, parity_expt}. This is made possible by an isomorphic correspondence between two seemingly different descriptions of the electromagnetic field: the single mode multiphoton Hilbert space expressed in the Fock state basis and the single photon multimode electromagnetic spanned by transverse spatial eigenmodes. Through this correspondence, operations acting on pseudospin degrees of freedom can be faithfully mapped onto experimentally accessible transformations and measurements in spatial parity space.


Consider a single photon state in the spatial domain, 
$|\Psi\rangle = \int dx\, \psi(x)\, |x\rangle,$ with $\quad 
\int dx\,|\psi(x)|^{2} = 1$.
This state is formally infinite-dimensional, the transverse profile \(\psi(x)\) can be expanded in an orthonormal functional basis \(\{\phi_{n}(x)\}\) as
$\psi(x) = \sum_{n} c_{n}\, \phi_{n}(x), \quad \text{with} \quad
|\Psi\rangle = \sum_{n} c_{n}\, |n\rangle$, which enables the association
$|n\rangle = \int dx\, \phi_{n}(x)\, |x\rangle $.

A suitable choice for the discrete orthonormal basis $\{\phi_n(x)\}$ is provided by the set of Hermite--Gaussian functions. A noteworthy property of this basis is the alternating parity of its elements: functions with even indices are even, while those with odd indices are odd. 
Mathematically, this can be expressed as
\[
\phi_{2n}(-x) = \phi_{2n}(x), \; 
\phi_{2n+1}(-x) = -\phi_{2n+1}(x), \; \forall\, n \in \mathbb{N}.
\]
A key observation is that the wavefunction can be decomposed into its even, {\it i.e.}, $\psi_{\mathrm{e}}(x) = \tfrac{1}{2}\big[\psi(x) + \psi(-x)\big]$ and odd, {\it i.e.}, $\psi_{\mathrm{o}}(x) = \tfrac{1}{2}\big[\psi(x) - \psi(-x)\big] $ parts, which are orthogonal,\(\int dx\, \psi_{\mathrm{e}}^{*}(x)\,\psi_{\mathrm{o}}(x)=0\). This naturally defines a two-dimensional ``parity qubit'' with logical basis states
\begin{equation}
|\mathrm{e}\rangle \propto \sum_{n} c_{2n}\, |2n\rangle, 
\quad
|\mathrm{o}\rangle \propto \sum_{n} c_{2n+1}\, |2n+1\rangle,
\end{equation}
so that the original state may be expressed in the 2D space of spatial parity as $|\Psi\rangle = e |\mathrm{e}\rangle + o |\mathrm{o}\rangle$,
where \(e\) and \(o\) are amplitudes. With this association, it is possible to decompose the state into even ($\ket{e}$) and odd ($\ket{o}$) spatial parity components.

In the spatial representation, the pseudospin operators ($q_i=0$) acting on $\psi(x)$ are defined as~\cite{spatial_parity}
\begin{align}
    \bar{s}_k = \iint dx dx' h_k (x,x') \ket{x}\bra{x'},\qquad k={1,2,3},\label{spatial_parity_operators}
\end{align}
where $h_k(x,x')$ is the impulse response function given by,
\begin{align}
    h_x(x,x') &= \sum_{n} [ \phi_{2n+1} (x) \phi_{2n}^{\star} (x') + \phi_{2n} (x) \phi_{2n+1}^{\star} (x')], \nonumber\\
    h_y(x,x') &= i \sum_{n} [ \phi_{2n+1} (x) \phi_{2n}^{\star} (x') - \phi_{2n} (x) \phi_{2n+1}^{\star} (x')], \nonumber\\
    h_z(x,x') &= \sum_{n} [ \phi_{2n} (x) \phi_{2n}^{\star} (x') - \phi_{2n+1} (x) \phi_{2n+1}^{\star} (x')].
\end{align}
In the basis of spatial eigenmodes, using  $|n\rangle = \int dx\, \phi_{n}(x)\, |x\rangle $, the spatial parity operators defined in Eq.~\eqref{spatial_parity_operators} can be directly identified with the pseudospin operators introduced in Eq.~\eqref{sxyz}. Acting with a spatial parity operator on the state $\ket{\Psi}$ yields
\begin{align}
    \bar{s}_k \ket{\Psi}
    &= \bar{s}_k \Big( \sum_n c_n \ket{n} \Big)
    = \sum_n c_n' \ket{n} \nonumber\\
    &= \int dx \, \psi'(x)\, \ket{x},
\end{align}
where the transformed spatial wave function is given by
\begin{equation}
    \psi'(x) = \int dx'\, h_k(x,x')\, \psi(x').
\end{equation}
This correspondence establishes a direct operational link between pseudospin transformations and experimentally accessible operations in spatial parity space, thereby enabling the implementation of pseudospin measurements through spatial parity observables.


The experimental elements~\cite{parity_expt} required for probing spatial parity consist of two basic components: a {\it parity rotator} and a {\it parity analyzer}. The {\it parity rotator} implements a rotation in the even--odd spatial parity subspace and is realized using a phase plate that introduces a relative phase $\theta$ between the $x \geq 0$ and $x < 0$ halves of the transverse spatial plane. This operation is formally equivalent to an $\mathrm{SO}(2)$ rotation acting on the effective two-dimensional parity space. The {\it parity analyzer} performs a projective measurement in the even--odd basis and is implemented using a parity sensitive Mach--Zehnder interferometer, in which a spatial flipper (a spatial flipper implements a transverse spatial inversion of the optical field, $\phi(x) \rightarrow \phi(-x)$
) is inserted in one interferometer arm, rendering the interference outcome parity dependent. 

A schematic of the bilocal measurement configuration is shown in Fig.~\ref{fig:bilocality_expt}. Two independent photon pair sources distribute entangled photons between Alice and Bob, and between Bob and Charlie, respectively. Alice applies a parity rotation characterized by the angle $\theta_1$, followed by a parity analyzer. Bob receives two photons, one from each source, and independently applies parity rotations characterized by the angle $\theta_2$ to both photons, followed by separate parity analyzers. Charlie applies a parity rotation characterized by the angle $\theta_3$, followed by a parity analyzer. Each parity analyzer yields a dichotomic outcome, where an even-parity detection event is assigned the value $+1$ and an odd-parity detection event is assigned the value $-1$.

The experimental data required for the bilocality test consist of coincidence counts corresponding to all possible combinations of parity outcomes. For each setting of the rotation angles $(\theta_1,\theta_2,\theta_3)$, one records coincidence counts
\[
N(a,b_1,b_2,c \,|\, \theta_1,\theta_2,\theta_3),
\]
where $a,b_1,b_2,c \in \{+1,-1\}$ denote the parity outcomes at Alice, Bob, and Charlie, respectively. For instance, the outcome $(+,+,+,-)$ corresponds to even-parity detections at Alice and Bob's detectors and an odd-parity detection at Charlie, while $(-,-,-,-)$ corresponds to odd-parity detections at all four detectors.

From these coincidence counts, the correlation function is defined as
\begin{equation}
E(\theta_1,\theta_2,\theta_3)
=
\frac{
\sum_{a,b_1,b_2,c=\pm 1} a b_1 b_2 c \,
N(a,b_1,b_2,c \,|\, \theta_1,\theta_2,\theta_3)
}{
\sum_{a,b_1,b_2,c=\pm 1}
N(a,b_1,b_2,c \,|\, \theta_1,\theta_2,\theta_3)
}.
\end{equation}

The bilocality inequality involves two nonlinear combinations of such correlators, denoted by $I$ and $J$, which are constructed from correlation functions evaluated at different measurement settings. Explicitly,
\begin{align}
I &= \sum_{x,z=0,1} E(\theta_1^{(x)},\theta_2,\theta_3^{(z)}),\nonumber\\
J &= \sum_{x,z=0,1} (-1)^{x+z}
E(\theta_1^{(x)},\theta_2,\theta_3^{(z)}),
\end{align}
where $\theta_1^{(x)}$ and $\theta_3^{(z)}$ denote two measurement settings for Alice and Charlie, respectively, while Bob performs fixed parity measurements. These quantities are then combined according to the bilocality inequality in Eq.~\eqref{bilocal}.

Since the spatial parity rotators and analyzers required for these measurements have already been implemented experimentally in Bell tests~\cite{parity_expt}, the above procedure provides a direct and experimentally grounded route for probing nonbilocal correlations using spatial parity observables.


\section{Conclusions}\label{s7}

Quantum networks provide a powerful setting for investigating nonlocal correlations that go beyond the conventional Bell framework. These networks facilitate the distribution of entanglement over long distances without requiring direct interaction between distant nodes, making them highly suitable for practical quantum architectures and repeaters~\cite{swap, repeater}. While the study of nonlocal correlations in DV quantum networks has seen substantial progress~\cite{Tavakoli2022}, CV networks have hitherto received very limitation attention. This is despite the notable advantages offered by CV systems—including near-deterministic state preparation, experimental accessibility, and their compatibility with long-distance communication~\cite{Braunstein2005}. In this context, our work takes a critical step toward understanding and leveraging nonlocal correlations in CV-based quantum networks.

In this work, we have proposed a formalism to probe CV network nonlocality using pseudospin measurements. Specifically, we have investigated here 
network nonlocality corrsponding to the scenarios of two basic open
network topologies, {\it viz.}, the linear chain network and the star
network. As a first application of the pseudospin framework in networks, we  have presented a study of the Gaussian two-mode squeezed vaccum state, showing reasonable persistence of n-locality violation even in the presence of thermal noise. 
We have further demonstrated network nonlocality for a wide range of CV resources, {\it i.e.,} several categories of non-Gaussian states such as  photon subtracted squeezed states,  entangled coherent states and  CV Werner-class states. Our results substantiate the role of non-Gaussianity as a powerful enabler of network nonlocality. Moreover, by relating pseudospin measurements to spatial parity observables, we have proposed an experimentally grounded framework for a practical demonstration of CV network nonlocality.

To summarize, we highlight some of the new and significant results on network nonlocality with CV systems, that have been revealed through our analysis  here. First, it is shown that the strength
of nonlocality in the star network configuration with TMSV states stays independent of the network size,  a result with potentially wide applicability in information processing protocols based on 
scaled-up quantum networks~\cite{kubala2026}. Another key result of our analysis is the robustness of CV network nonlocality against local thermal noise: we have shown that nonbilocal correlations can persist even at arbitrarily high temperatures, provided the squeezing exceeds a critical threshold. Next, we have established non-Gaussianity to enhance network nonlocality beyond the Gaussian paradigm, with specific illustrations across several classes of non-Gaussian states. In particular, we have found that a coherent superposition of single photon subtractions across modes yields maximal bilocality violation even in the absence of squeezing, a regime of direct experimental relevance.

Our findings contribute significantly to the understanding of network nonlocality in infinite-dimensional systems and open up several promising avenues for future research. Though we have studied the effects of thermal noise in this work, the impact of other kinds of environmental effects
on the robustness of network nonlocality merit further exploration. 
Another interesting direction would be  systematic investigations of the role of non-Gaussian operations on network nonlocality, and their interplay with resources such as entanglement and squeezing. Moreover, our approach
motivates extension to other types of open and possibly closed network
configurations. Finally, adaptation  of our analysis to measurement schemes involving genuine continuous outcomes might offer deeper insights into the structure of nonlocal correlations in the CV regime.

\appendix
\section{Network nonlocality with non-Gaussian states}\label{s8}

\subsection{Photon subtracted squeezed vacuum state} \label{s8A}

\subsubsection{Asymmetric Photon Subtraction: Single Mode Operation}

Applying the annihilation operator \((\mathbb{I} \otimes \hat{b})\) on the second mode of a TMSV yields the unnormalized state
\begin{align}
\ket{\xi_1} &\propto \sum_{n=1}^{\infty} \lambda^n \sqrt{n} \ket{n, n-1},
\end{align}
where \(\lambda = \tanh r\) and \(\hat{b}\) is the annihilation operator for the second mode. Upon normalization and relabeling, this becomes
\begin{align}
\ket{\xi_1} &= (1 - \lambda^2) \sum_{n=0}^{\infty} \lambda^n \sqrt{n+1} \ket{n+1, n}.
\end{align}

For the state $\ket{\xi_1}$ the optimal pseudospin measurement settings correspond to the parameters $q=1$ for the first mode and $q=0$ for the second mode.

\vspace{2mm}
\textbf{\textit{Configuration A1:}} In the first scenario, Alice and Bob share the photon subtracted state \(\ket{\xi_1}\), while Bob and Charlie share the standard TMSV state \(\ket{\zeta}\). That is,
\begin{align*}
\rho_{AB} = \ket{\xi_1}\bra{\xi_1}, \qquad \rho_{BC} = \ket{\zeta}\bra{\zeta}.
\end{align*}
The two largest eigenvalues of the correlation matrices \(R^{AB}\) and \(R^{BC}\) (see Sec.~\ref{s6}), in this case are found to be
\begin{align}
    \nu_1^{(a)} &= 1, \qquad \nu_2^{(a)} = 2 \lambda (1 - \lambda^2)^2 \sum_{n=0}^{\infty} \lambda^{4n} \sqrt{(2n+1)(2n+2)}, \\
    \mu_1 &= 1, \qquad \mu_2 = \frac{2\lambda}{1 + \lambda^2} = K(r).
\end{align}
Thus, the maximal value of the bilocality expression becomes
\begin{align}
S^{\max}_{\xi_1,\zeta} = 2 \sqrt{\nu_1^{(a)} \mu_1 + \nu_2^{(a)} \mu_2}.
\end{align}

\vspace{2mm}
\textbf{\textit{Configuration A2:}} In the second configuration, both shared states are identical and given by the photon subtracted state \(\ket{\xi_1}\), i.e.,
\begin{align*}
\rho_{AB} = \rho_{BC} = \ket{\xi_1}\bra{\xi_1}.
\end{align*}
In this case, the maximal value of the bilocality expression is,
\begin{align}
S^{\max}_{\xi_1,\xi_1} = 2 \sqrt{ (\nu_1^{(a)})^2 + (\nu_2^{(a)})^2 }.
\end{align}

\vspace{2mm}
\noindent

\subsubsection{Symmetric Photon Subtraction: Dual-Mode Operation}

Applying the annihilation operators \((\hat{a} \otimes \hat{b})\) on the TMSV yields the normalized two-mode state
\begin{align}
    \ket{\xi_2} &= \left( \frac{1-\lambda^2}{\lambda \sqrt{1+\lambda^2}} \right) \sqrt{1 - \lambda^2} \sum_{n=1}^{\infty} \lambda^n n \ket{n-1, n-1} \nonumber \\
                &= (1 - \lambda^2) \sqrt{\frac{1-\lambda^2}{1+\lambda^2}} \sum_{n=0}^{\infty} \lambda^n (n+1) \ket{n, n},
\end{align}
where \(\lambda = \tanh r\), and \(\hat{a}\), \(\hat{b}\) denote annihilation operators for the first and second mode, respectively.

For the state $\ket{\xi_2}$ the optimal pseudospin measurement settings correspond to the parameters $q_i=0$ for both the modes.

\vspace{2mm}
\textbf{\textit{Configuration B1:}} In this configuration, the entangled state shared between Alice and Bob is \(\ket{\xi_2}\), while Bob and Charlie share the original TMSV state \(\ket{\zeta}\):
\begin{align*}
\rho_{AB} = \ket{\xi_2}\bra{\xi_2}, \qquad \rho_{BC} = \ket{\zeta}\bra{\zeta}.
\end{align*}
The two largest eigenvalues of the corresponding correlation matrix \(R^{AB}\) are given by
\begin{align}
    \nu_1^{(s)} = 1, \qquad \nu_2^{(s)} = \frac{2 \lambda (1-\lambda^2)^3 (6\lambda^4+2)}{(1+\lambda^2)(1-\lambda^4)^3},
\end{align}
The resulting maximal bilocality expression becomes
\begin{align}
S^{\max}_{\xi_2,\zeta} = 2 \sqrt{\nu_1^{(s)} \mu_1 + \nu_2^{(s)} \mu_2}.
\end{align}

\vspace{2mm}
\textbf{\textit{Configuration B2:}} Here, both shared states are taken to be \(\ket{\xi_2}\), making this a fully symmetric setting:
\begin{align*}
\rho_{AB} = \rho_{BC} = \ket{\xi_2}\bra{\xi_2}.
\end{align*}
The maximum bilocality expression in this configuration becomes
\begin{align}
S^{\max}_{\xi_2,\xi_2} = 2 \sqrt{ (\nu_1^{(s)})^2 + (\nu_2^{(s)})^2 }.
\end{align}


\subsubsection{Coherent Superposition of Photon Subtractions across Modes} 

 The state is represented as \cite{chowdhury2014, mallick2025}, 
\begin{equation}
\begin{split}
    \ket{\xi_{3}}= & \frac{1}{\sqrt{2}\sinh{r}}\sqrt{1-\lambda^2}\sum_{n=1}^{\infty}\lambda^n\sqrt{n} [ \ket{n-1,n} \\ & +(-1)^k\ket{n,n-1} ] \\
    = & \frac{1}{\sqrt{2}}(1-\lambda^2)\sum_{n=0}^{\infty}\lambda^n\sqrt{n+1} [ \ket{n,n+1} \\ & +(-1)^k\ket{n+1,n} ]
    \end{split}
\end{equation}
with $\lambda = \tanh{r}, \hspace{0.1cm} r >0$. Here we consider the state with $k=0$.
In order to reveal the nonbilocal nature of the state \(\ket{\xi_3}\), we evaluate the bilocality expression using pseudospin observables with \(q_i = 0\).

\vspace{2mm}
\textbf{\textit{Configuration C1:}} In this configuration, the entangled state shared between Alice and Bob is \(\ket{\xi_3}\), while Bob and Charlie share the original TMSV state \(\ket{\zeta}\):
\begin{align*}
\rho_{AB} = \ket{\xi_3}\bra{\xi_3}, \qquad \rho_{BC} = \ket{\zeta}\bra{\zeta}.
\end{align*}
The two largest eigenvalues of the corresponding correlation matrix \(R^{BC}\) are given by
\begin{align}
    \nu_1^{(c)} &= 1, \qquad \nu_2^{(c)} = \frac{1+\lambda^4}{(1+\lambda^2)^2}.
\end{align}
The resulting maximal bilocality expression becomes
\begin{align}
S^{\max}_{\xi_3,\zeta} = 2 \sqrt{\nu_1^{(s)} \mu_1 + \nu_2^{(s)} \mu_2}.
\end{align}

\vspace{2mm}
\textbf{\textit{Configuration C2:}} Here, both shared states are taken to be \(\ket{\xi_2}\), making this a fully symmetric setting:
\begin{align*}
\rho_{AB} = \rho_{BC} = \ket{\xi_3}\bra{\xi_3}.
\end{align*}
The maximum bilocality expression in this configuration becomes

\begin{align}
S^{\max}_{\xi_3,\xi_3} = 2 \sqrt{ (\nu_1^{(c)})^2 + (\nu_2^{(c)})^2 }.
\end{align}

\subsection{Entangled coherent state (ECS)} \label{s8B}
The entangled coherent state (ECS) can be defined as\cite{dichotomic_nonlocality},
\begin{eqnarray}
|{\rm ECS}\rangle_\alpha ={\cal N}(|\alpha\rangle|-\alpha\rangle
 -|-\alpha\rangle|\alpha\rangle), \;\;\;
\end{eqnarray}
where, $|\alpha\rangle =
e^{-|\alpha|^2/2}\sum_{n=0}^{\infty}\frac{\alpha^n}{\sqrt{n!}}|n\rangle$, $|\alpha\rangle$ is a coherent state with $\alpha\neq0$, we have considered $\alpha$ to be real for simplicity. ${\cal N}$ is a normalization factor given by,
\begin{equation}
   { \cal {N}} = \frac{1}{\sqrt{2(1-e^{-4\alpha^2})}}
\end{equation}

We consider the bilocal network scenario where the states shared between Alice and Bob $(\ket{\psi_{AB}})$, and the state shared between Bob and Charlie $(\ket{\psi_{BC}}$ both are two entangled coherent states, $\ket{\rm{ECS}}_\alpha$ and $\ket{\rm{ECS}}_\beta$ respectively. Using Eq.~(\ref{I}) and Eq.~(\ref{J}), we get,
\begin{align}
    \langle A_x B_0 C_z\rangle_{| {\rm{ECS}} \rangle_{\alpha} \otimes |{\rm{ECS}}\rangle_{\beta}} &=  \cos(\theta_1)\cos(\theta_2)
    \\ \langle A_x B_1 C_z\rangle_{|{\rm{ECS}} \rangle_{\alpha} \otimes |{\rm{ECS}}\rangle_{\beta}} &=  Q^2(\alpha,\beta) \sin(\theta_1) \sin(\theta_2)
\end{align}
where,
\begin{align}
Q^2(\alpha,\beta) &=   \frac{\left( \sum\limits_{n,m =0}^\infty \frac{\alpha^{4n+1}}{\sqrt{(2n)! (2n+1)!}} \frac{\beta^{4m+1}}{\sqrt{(2m)! (2m+1)!}}\right)^2}{\cosh (\alpha^2) \sinh (\alpha^2)\cosh (\beta^2) \sinh (\beta^2)}
\end{align}
Using Eq.~(\ref{IJ1}), and Eq.~ (\ref{IJ2}), we can calculate the values of $I$, $J$ and $S$,
\begin{align}
    I_{\alpha, \beta} &= 4 \cos(\theta_1)\cos(\theta_2) \nonumber\\
    J_{\alpha, \beta} &=  4 Q^2(\alpha,\beta) \sin(\theta_1) \sin(\theta_2) \nonumber\\
    S_{\alpha, \beta} &= 2 \left[ \sqrt{\cos(\theta_1)\cos(\theta_2)} +Q(\alpha,\beta) \sqrt{\sin(\theta_1) \sin(\theta_2)}  \right] 
\end{align}
The optimal measurement settings, which maximize the value of $S_{\zeta}$, also the violation of the bilocality inequality considered in Eq.~(\ref{bilocal}), are given by,
\begin{align}
& \theta_1 = \theta_2 = \theta = \arctan (Q)\\
 \text{Hence,}\hspace{2mm}  \cos & \theta = \frac{1}{\sqrt{1+Q^2}} \hspace{2mm} \text{and} \hspace{2mm} \sin{\theta} = \frac{Q}{\sqrt{1+Q^2}} \nonumber
\end{align}
such that the maximal bilocality quantity $S^{\max}$ reduces to,
\begin{align}
    S^{\max}_{\alpha, \beta} = 2\sqrt{1+Q^2}
\end{align}

\subsection{CV Werner state} \label{s8C}

The CV Werner state takes the form
\begin{equation}
\rho_W = p\,\rho_{\text{TMSV}} + (1-p)\,\rho_T,\quad 0 \leq p \leq 1,
\end{equation}
where $\rho_{\text{TMSV}} = \ket{\zeta}\bra{\zeta}$, and $\ket{\zeta}$ is the TMSV state (\ref{nopa}). This defines a class of states parametrized by $(p, r, s)$. For simplicity, we take $r = s$, reducing to two parameters $(p, r)$.

We now consider the bilocal network scenario where the states shared between Alice and Bob $(\ket{\psi_{AB}})$, and the state shared between Bob and Charlie $(\ket{\psi_{BC}}$ both are CV Werner states with parameters $(p_1 , r)$ and $(p_2, r)$ respectively.  Using Eqs.~(\ref{I}) and (\ref{J}), we get,
\begin{align}
    \langle A_x B_0 C_z\rangle_{\rho_{W_1} \otimes \rho_{W_2}} &=  T_I \cos(\theta_1)\cos(\theta_2)
    \\ \langle A_x B_1 C_z\rangle_{\rho_{W_1} \otimes \rho_{W_2}} &=  T_J \sin(\theta_1) \sin(\theta_2)
\end{align}
where 
\begin{align}
    T_I &= p_1 p_2 + (1-K^2)(p_1 - 2p_1p_2+p_2) \nonumber \\
    &+ (1-p_1)(1-p_2)(1-K^2)^2 \; , \; \nonumber\\
    T_J &= p_1p_2K^2 \;, 
\end{align}
with $K = \tanh {2r}$. Using Eq.~(\ref{IJ1}), and Eq.~(\ref{IJ2}), we can calculate the values of $I$, $J$ and $S$,
\begin{align}
    I_{\rho_{W_1} \otimes \rho_{W_2}} &= 4 T_I \cos(\theta_1)\cos(\theta_2) \nonumber\\
    J_{\rho_{W_1} \otimes \rho_{W_2}} &=  4 T_J\sin(\theta_1) \sin(\theta_2) \nonumber\\
    S_{\rho_{W_1} \otimes \rho_{W_2}} &= 2 \left[ \sqrt{T_I \cos(\theta_1)\cos(\theta_2)} + \sqrt{ T_J \sin(\theta_1) \sin(\theta_2)}  \right] 
\end{align}
 
The optimal measurement settings which maximize the violation of the bilocality inequality considered in Eq.~(\ref{bilocal}), are given by,
\begin{align}
& \theta_1 = \theta_2 = \theta = \arctan \left (\sqrt{\frac{T_J}{T_I}} \; \; \right)\\
\end{align}
such that the maximal bilocality quantity $S^{\max}$ reduces to,
\begin{align}
    S^{\max}_{{\rho_{W_1} , \; \rho_{W_2}}} = 2\sqrt{T_I+T_J}
\end{align}

\section{Derivations Supporting the Observations}\label{s9}

\subsection{Derivations supporting Observation~\ref{temp}}\label{s9_obs4}

The bilocality inequality violation condition for the noisy TMSV state can be expressed as
\begin{equation}
\mathcal{Q}(\epsilon)=\mathcal{D}p^2+\mathcal{E}p+\mathcal{F} > 0 ,
\label{eq:app_start}
\end{equation}
with
\begin{align}
\mathcal{D} &= K^2+(1-\epsilon)^2, \\
\mathcal{E} &= 2\epsilon(1-\epsilon), \\
\mathcal{F} &= \epsilon^2-1,
\end{align}
and
\begin{equation}
\epsilon=\tanh\!\left(\frac{\beta_1}{2}\right)
\tanh\!\left(\frac{\beta_2}{2}\right).
\end{equation}
Since $\beta_i = 1/T_i$ and $T_i>0$, we have
\begin{equation}
0 \le \epsilon \le 1 .
\label{eq:eps_domain}
\end{equation}

Substituting the coefficients into Eq.~(\ref{eq:app_start}) yields
\begin{align}
& \left[K^2+(1-\epsilon)^2\right]p^2 + 2\epsilon(1-\epsilon)p + \epsilon^2-1 & > 0 \nonumber \\
\Rightarrow \quad&(1-p)^2\epsilon^2 + 2p(1-p)\epsilon + p^2(K^2+1)-1 &> 0 \label{eq:Qfinal}
\end{align}

Thus, $\mathcal{Q}(\epsilon)$ is convex in $\epsilon$, since $(1-p)^2 \ge 0$.
The stationary point is determined by imposing
\(
{\partial \mathcal{Q}(\epsilon)}/{\partial \epsilon}=0,
\)

\begin{equation}
\epsilon_*=-\frac{p}{1-p},
\end{equation}
which lies outside the physical domain (\ref{eq:eps_domain}). Therefore
the minimum of $\mathcal{Q}(\epsilon)$ within the accessible parameter range
occurs at the boundary $\epsilon=0$, when temperature of either one or both noise is infinity. Evaluating Eq.~(\ref{eq:Qfinal}),
\begin{equation}
\mathcal{Q}_{\min}=\mathcal{Q}(0)=p^2(K^2+1)-1 .
\end{equation}
A temperature-independent sufficient condition ensuring
$\mathcal{Q}(\epsilon)>0$ for all $\beta_1,\beta_2$ is therefore
\begin{align}
p^2(K^2+1) -1 > 0 \quad \Rightarrow \quad K>\sqrt{\frac{1}{p^2}-1},
\end{align}
which establishes Observation~\ref{temp}.

\subsection{Derivation supporting Observation~\ref{B2vsA2}}\label{s9_obs6}



To compare the strength of bilocality violations produced by symmetric
and asymmetric photon subtraction, we note that for all configurations
considered here the maximal bilocality expressions depend
monotonically on the second singular value of the corresponding
correlation matrix.

For configurations A1 and B1 one has
\begin{align}
S^{\max}_{\xi_1,\zeta}
&=2\sqrt{1+\nu_2^{(a)}(\lambda)\mu_2(\lambda)},\\
S^{\max}_{\xi_2,\zeta}
&=2\sqrt{1+\nu_2^{(s)}(\lambda)\mu_2(\lambda)},
\end{align}
while for configurations A2 and B2,
\begin{align}
S^{\max}_{\xi_1,\xi_1}
&=2\sqrt{1+\big(\nu_2^{(a)}(\lambda)\big)^2},\\
S^{\max}_{\xi_2,\xi_2}
&=2\sqrt{1+\big(\nu_2^{(s)}(\lambda)\big)^2}.
\end{align}
Since $\mu_2(\lambda)>0$ and the square-root function is monotonic,
both comparisons reduce to determining the sign of
\begin{equation}
\Delta (\lambda) = \nu_2^{(s)}(\lambda) - \nu_2^{(a)}(\lambda).
\label{eq:keycomparison}
\end{equation}

Substituting
\begin{align*}
\nu_2^{(a)}(\lambda)
&=
2\lambda(1-\lambda^2)^2
\sum_{n=0}^{\infty}\lambda^{4n}\sqrt{(2n+1)(2n+2)},\\
\nu_2^{(s)}(\lambda)
&=
\frac{2\lambda(1-\lambda^2)^3(6\lambda^4+2)}
{(1+\lambda^2)(1-\lambda^4)^3},
\end{align*}
we evaluate their difference numerically. An exact analytical evaluation of \(\nu_2^{(a)} (\lambda)\) is hindered by the square-root structure within the summation. To estimate it accurately, we use a numerical truncation technique: the summation is carried out up to \(n = N\) terms, and convergence is confirmed if the difference between partial sums with \(N\) and \(N+1\) terms falls below \(10^{-20}\). Once this criterion is met, we consider the sum to have converged with sufficient precision. Computing $\nu_2^{(a)} (\lambda)$ to convergence and solving $\Delta(\lambda)=0$ yields a single crossing point
\begin{equation}
\lambda_c \approx 0.616 .
\end{equation}
We find
\begin{equation}
\nu_2^{(s)}(\lambda)>\nu_2^{(a)}(\lambda)
\quad \text{for} \quad
0\le\lambda\le\lambda_c,
\end{equation}
with the ordering reversed beyond this value.

Consequently, symmetric photon subtraction produces a larger maximal
bilocality violation than the asymmetric scheme for both A1 vs B1 and
A2 vs B2 whenever $0 < \lambda < 0.616$, establishing
Observation~\ref{B2vsA2}.

\subsection{Derivation supporting Observation~\ref{c1_enhancement}}\label{s9_obs8}

Observation ~\ref{c1_enhancement} states that configurations C1 and C2 exhibit enhancement in violation over TMSV up in the squeezing parameter range, \(0< \lambda < 0.4354\). To justify the observation, we compare the maximal violations with that of the
TMSV case. For TMSV states the two largest singular values are
$\mu_1=1$ and $\mu_2=K(r)=\frac{2\lambda}{1+\lambda^2}$,
so that
\begin{equation}
\mathcal{S}_{\mathrm{bilocal\_TMSV}}^{\max}
=2\sqrt{1+\left({\frac{2\lambda}{1+\lambda^2}}\right)^2}.
\end{equation}

For the photon-subtracted configurations we have
\begin{equation}
\nu_1^{(c)}=1, \qquad
\nu_2^{(c)}=\frac{1+\lambda^4}{(1+\lambda^2)^2}.
\end{equation}
Configurations C1 and C2 yields,
\begin{align}
S^{\max}_{\xi_3,\zeta}
&=2\sqrt{1+\left(\frac{1+\lambda^4}{(1+\lambda^2)^2}\right)\left( \frac{2\lambda}{1+\lambda^2}\right)}, \quad \\
S^{\max}_{\xi_3,\xi_3}
&=2\sqrt{1+\left(\frac{1+\lambda^4}{(1+\lambda^2)^2}\right)^2}.
\end{align}
Hence enhancement over the TMSV value occurs whenever
\begin{align}
    &\frac{1+\lambda^4}{(1+\lambda^2)^2} > \frac{2\lambda}{1+\lambda^2} \nonumber \\
    \Rightarrow \quad &\lambda^4 - 2\lambda^3 - 2\lambda + 1  = \mathcal{L}(\lambda)> 0
\end{align} 
The boundary between enhancement and no enhancement is determined by
\begin{equation}
\mathcal{L} (\lambda) = \lambda^4 - 2\lambda^3 - 2\lambda + 1 = 0 .
\end{equation}
The only physically relevant root in the interval $0<\lambda<1$ is
\begin{equation}
\lambda_* = \frac{1+\sqrt{3}-12^{1/4}}{2} \approx 0.4354.
\end{equation}
$\mathcal{L}(\lambda)$ is positive in the range $0 < \lambda < \lambda_*$, and negative in the range $\lambda_* < \lambda < 1$. Hence configurations C1 and C2 exhibit an advantage over the TMSV resource
whenever
\begin{equation}
0 < \lambda < \lambda_* \approx 0.4354.
\end{equation}

Thus, within this squeezing range the maximal bilocal violation produced by the
coherent photon-subtracted resources exceeds that of the TMSV state, establishing the observation.

\bibliography{main}

@article{Bellpaper,
  title = {On the Einstein Podolsky Rosen paradox},
  author = {Bell, J. S.},
  journal = {Physics Physique Fizika},
  volume = {1},
  issue = {3},
  pages = {195--200},
  numpages = {6},
  year = {1964},
  month = {Nov},
  publisher = {American Physical Society},
  doi = {10.1103/PhysicsPhysiqueFizika.1.195},
  url = {https://link.aps.org/doi/10.1103/PhysicsPhysiqueFizika.1.195}
}

@article{epr,
  title = {Can Quantum-Mechanical Description of Physical Reality Be Considered Complete?},
  author = {Einstein, A. and Podolsky, B. and Rosen, N.},
  journal = {Phys. Rev.},
  volume = {47},
  issue = {10},
  pages = {777--780},
  numpages = {0},
  year = {1935},
  month = {May},
  publisher = {American Physical Society},
  doi = {10.1103/PhysRev.47.777},
  url = {https://link.aps.org/doi/10.1103/PhysRev.47.777}
}

@article{Rev1,
  title = {Nonlocality and communication complexity},
  author = {Buhrman, Harry and Cleve, Richard and Massar, Serge and de Wolf, Ronald},
  journal = {Rev. Mod. Phys.},
  volume = {82},
  issue = {1},
  pages = {665--698},
  numpages = {0},
  year = {2010},
  month = {Mar},
  publisher = {American Physical Society},
  doi = {10.1103/RevModPhys.82.665},
  url = {https://link.aps.org/doi/10.1103/RevModPhys.82.665}
}

@article{Rev2,
  title = {Bell nonlocality},
  author = {Brunner, Nicolas and Cavalcanti, Daniel and Pironio, Stefano and Scarani, Valerio and Wehner, Stephanie},
  journal = {Rev. Mod. Phys.},
  volume = {86},
  issue = {2},
  pages = {419--478},
  numpages = {60},
  year = {2014},
  month = {Apr},
  publisher = {American Physical Society},
  doi = {10.1103/RevModPhys.86.419},
  url = {https://link.aps.org/doi/10.1103/RevModPhys.86.419}
}

@incollection{mckague,
  author    = {M. McKague and M. Mosca},
  title     = {Generalized self-testing and the security of the 6-state protocol},
  booktitle = {Theory of Quantum Computation, Communication, and Cryptography},
  year      = {2011},
  volume    = {6519},
  isbn      = {978-3-642-18072-9},
  doi       = {10.1007/978-3-642-18073-6_10},
  publisher = {Springer},
  url       = {https://doi.org/10.1007/978-3-642-18073-6_10}
}

@article{mckague12,
  author  = {M. McKague and T. H. Yang and V. Scarani},
  title   = {Robust Self Testing of the Singlet},
  journal = {J. Phys. A: Math. Theor.},
  volume  = {45},
  number  = {45},
  pages   = {455304},
  year    = {2012},
  doi     = {10.1088/1751-8113/45/45/455304},
  url     = {https://doi.org/10.1088/1751-8113/45/45/455304}
}

@article{supicrev,
  author  = {I. Supic and J. Bowles},
  title   = {Self-testing of quantum systems: A review},
  journal = {Quantum},
  volume  = {4},
  pages   = {337},
  year    = {2020},
  doi     = {10.22331/q-2020-09-30-337},
  url     = {https://doi.org/10.22331/q-2020-09-30-337}
}

@article{swap,
  title = {``Event-ready-detectors'' Bell experiment via entanglement swapping},
  author = {\ifmmode \dot{Z}\else \.{Z}\fi{}ukowski, M. and Zeilinger, A. and Horne, M. A. and Ekert, A. K.},
  journal = {Phys. Rev. Lett.},
  volume = {71},
  issue = {26},
  pages = {4287--4290},
  numpages = {0},
  year = {1993},
  month = {Dec},
  publisher = {American Physical Society},
  doi = {10.1103/PhysRevLett.71.4287},
  url = {https://link.aps.org/doi/10.1103/PhysRevLett.71.4287}
}

@article{acin07,
  author  = {A. Acin and N. Brunner and N. Gisin and S. Massar and S. Pironio and V. Scarani},
  title   = {Device-Independent Security of Quantum Cryptography against Collective Attacks},
  journal = {Phys. Rev. Lett.},
  volume  = {98},
  pages   = {230501},
  year    = {2007},
  doi     = {10.1103/PhysRevLett.98.230501},
  url     = {https://journals.aps.org/prl/abstract/10.1103/PhysRevLett.98.230501}
}

@article{Farkas2024,
  author  = {M. Farkas},
  title   = {Unbounded Device-Independent Quantum Key Rates from Arbitrarily Small Nonlocality},
  journal = {Phys. Rev. Lett.},
  volume  = {132},
  pages   = {210803},
  year    = {2024},
  doi     = {10.1103/PhysRevLett.132.210803},
  url     = {https://journals.aps.org/prl/abstract/10.1103/PhysRevLett.132.210803}
}

@article{QCrev,
  author  = {C. Portmann and R. Renner},
  title   = {Security in quantum cryptography},
  journal = {Rev. Mod. Phys.},
  volume  = {94},
  pages   = {025008},
  year    = {2022},
  doi     = {10.1103/RevModPhys.94.025008},
  url     = {https://journals.aps.org/rmp/abstract/10.1103/RevModPhys.94.025008}
}

@article{Gras2021,
  author  = {F. Grasselli and G. Murta and H. Kampermann and D. Bruß},
  title   = {Entropy Bounds for Multiparty Device-Independent Cryptography},
  journal = {PRX Quantum},
  volume  = {2},
  pages   = {010308},
  year    = {2021},
  doi     = {10.1103/PRXQuantum.2.010308},
  url     = {https://journals.aps.org/prxquantum/abstract/10.1103/PRXQuantum.2.010308}
}

@article{Piro2010,
  author  = {S. Pironio and others},
  title   = {Random numbers certified by Bell’s theorem},
  journal = {Nature},
  volume  = {464},
  pages   = {1021--1024},
  year    = {2010},
  doi     = {10.1038/nature09008},
  url     = {https://doi.org/10.1038/nature09008}
}

@article{Acin2012,
  author  = {A. Acín and S. Massar and S. Pironio},
  title   = {Randomness versus Nonlocality and Entanglement},
  journal = {Phys. Rev. Lett.},
  volume  = {108},
  pages   = {100402},
  year    = {2012},
  doi     = {10.1103/PhysRevLett.108.100402},
  url     = {https://journals.aps.org/prl/abstract/10.1103/PhysRevLett.108.100402}
}

@article{3gisin17,
  title = {Violation of Bell's Inequality over 4 km of Optical Fiber},
  author = {Tapster, P. R. and Rarity, J. G. and Owens, P. C. M.},
  journal = {Phys. Rev. Lett.},
  volume = {73},
  issue = {14},
  pages = {1923--1926},
  numpages = {0},
  year = {1994},
  month = {Oct},
  publisher = {American Physical Society},
  doi = {10.1103/PhysRevLett.73.1923},
  url = {https://link.aps.org/doi/10.1103/PhysRevLett.73.1923}
}

@article{4gisin17,
  title = {Violation of Bell Inequalities by Photons More Than 10 km Apart},
  author = {Tittel, W. and Brendel, J. and Zbinden, H. and Gisin, N.},
  journal = {Phys. Rev. Lett.},
  volume = {81},
  issue = {17},
  pages = {3563--3566},
  numpages = {0},
  year = {1998},
  month = {Oct},
  publisher = {American Physical Society},
  doi = {10.1103/PhysRevLett.81.3563},
  url = {https://link.aps.org/doi/10.1103/PhysRevLett.81.3563}
}

@article{kunduBi,
  title = {Maximal qubit violation of $n$-local inequalities in a quantum network},
  author = {Kundu, Amit and Molla, Mostak Kamal and Chattopadhyay, Indrani and Sarkar, Debasis},
  journal = {Phys. Rev. A},
  volume = {102},
  issue = {5},
  pages = {052222},
  numpages = {6},
  year = {2020},
  month = {Nov},
  publisher = {American Physical Society},
  doi = {10.1103/PhysRevA.102.052222},
  url = {https://link.aps.org/doi/10.1103/PhysRevA.102.052222}
}

@article{Bithree,
  author    = {Kaushiki Mukherjee and Biswajit Paul and Debasis Sarkar},
  title     = {Correlations in $n$-local scenario},
  journal   = {Quantum Information Processing},
  volume    = {14},
  number    = {6},
  pages     = {2025--2042},
  year      = {2015},
  month     = {June},
  doi       = {10.1007/s11128-015-0971-7},
  url       = {https://doi.org/10.1007/s11128-015-0971-7},
  issn      = {1573-1332}
}

@article{repeater2,
author = {Stephanie Wehner  and David Elkouss  and Ronald Hanson },
title = {Quantum internet: A vision for the road ahead},
journal = {Science},
volume = {362},
number = {6412},
pages = {eaam9288},
year = {2018},
doi = {10.1126/science.aam9288},
URL = {https://www.science.org/doi/abs/10.1126/science.aam9288},
eprint = {https://www.science.org/doi/pdf/10.1126/science.aam9288},
}

@article{repeater,
  title = {Quantum repeaters: From quantum networks to the quantum internet},
  author = {Azuma, Koji and Economou, Sophia E. and Elkouss, David and Hilaire, Paul and Jiang, Liang and Lo, Hoi-Kwong and Tzitrin, Ilan},
  journal = {Rev. Mod. Phys.},
  volume = {95},
  issue = {4},
  pages = {045006},
  numpages = {66},
  year = {2023},
  month = {Dec},
  publisher = {American Physical Society},
  doi = {10.1103/RevModPhys.95.045006},
  url = {https://link.aps.org/doi/10.1103/RevModPhys.95.045006}
}

@article{gisin_pure,
  title = {All entangled pure quantum states violate the bilocality inequality},
  author = {Gisin, Nicolas and Mei, Quanxin and Tavakoli, Armin and Renou, Marc Olivier and Brunner, Nicolas},
  journal = {Phys. Rev. A},
  volume = {96},
  issue = {2},
  pages = {020304},
  numpages = {5},
  year = {2017},
  month = {Aug},
  publisher = {American Physical Society},
  doi = {10.1103/PhysRevA.96.020304},
  url = {https://link.aps.org/doi/10.1103/PhysRevA.96.020304}
}

@article{CV_Werner,
  title = {Continuous-variable Werner state: Separability, nonlocality, squeezing, and teleportation},
  author = {Mi\ifmmode \check{s}\else \v{s}\fi{}ta, Ladislav and Filip, Radim and Fiur\'a\ifmmode \check{s}\else \v{s}\fi{}ek, Jarom\'{\i}r},
  journal = {Phys. Rev. A},
  volume = {65},
  issue = {6},
  pages = {062315},
  numpages = {8},
  year = {2002},
  month = {Jun},
  publisher = {American Physical Society},
  doi = {10.1103/PhysRevA.65.062315},
  url = {https://link.aps.org/doi/10.1103/PhysRevA.65.062315}
}

@article{Chen_Maximal_Bell,
  title = {Maximal Violation of Bell's Inequalities for Continuous Variable Systems},
  author = {Chen, Zeng-Bing and Pan, Jian-Wei and Hou, Guang and Zhang, Yong-De},
  journal = {Phys. Rev. Lett.},
  volume = {88},
  issue = {4},
  pages = {040406},
  numpages = {4},
  year = {2002},
  month = {Jan},
  publisher = {American Physical Society},
  doi = {10.1103/PhysRevLett.88.040406},
  url = {https://link.aps.org/doi/10.1103/PhysRevLett.88.040406}
}

@article{GHZ_cv,
  title = {Greenberger-Horne-Zeilinger nonlocality for continuous-variable systems},
  author = {Chen, Zeng-Bing and Zhang, Yong-De},
  journal = {Phys. Rev. A},
  volume = {65},
  issue = {4},
  pages = {044102},
  numpages = {4},
  year = {2002},
  month = {Apr},
  publisher = {American Physical Society},
  doi = {10.1103/PhysRevA.65.044102},
  url = {https://link.aps.org/doi/10.1103/PhysRevA.65.044102}
}

@article{Horodecki_Bell,
    title = {Violating Bell inequality by mixed spin-12 states: necessary and sufficient condition},
    author = {R. Horodecki and P. Horodecki and M. Horodecki},
    journal = {Physics Letters A},
    volume = {200},
    number = {5},
    pages = {340-344},
    year = {1995},
    issn = {0375-9601},
    doi = {https://doi.org/10.1016/0375-9601(95)00214-N},
    url = {https://www.sciencedirect.com/science/article/pii/037596019500214N}
}

@article{photon_subtracted,
  title = {Response in the violation of the Bell inequality to imperfect photon addition and subtraction in noisy squeezed states of light},
  author = {Roy, Saptarshi and Chanda, Titas and Das, Tamoghna and Sen(De), Aditi and Sen, Ujjwal},
  journal = {Phys. Rev. A},
  volume = {98},
  issue = {5},
  pages = {052131},
  numpages = {16},
  year = {2018},
  month = {Nov},
  publisher = {American Physical Society},
  doi = {10.1103/PhysRevA.98.052131},
  url = {https://link.aps.org/doi/10.1103/PhysRevA.98.052131}
}

@article{Braunstein2005,
  title = {Quantum information with continuous variables},
  author = {Braunstein, Samuel L. and van Loock, Peter},
  journal = {Rev. Mod. Phys.},
  volume = {77},
  issue = {2},
  pages = {513--577},
  numpages = {0},
  year = {2005},
  month = {Jun},
  publisher = {American Physical Society},
  doi = {10.1103/RevModPhys.77.513},
  url = {https://link.aps.org/doi/10.1103/RevModPhys.77.513}
}

@article{Adesso2014,
author = {Adesso, Gerardo and Ragy, Sammy and Lee, Antony R.},
title = {Continuous Variable Quantum Information: Gaussian States and Beyond},
journal = {Open Systems \& Information Dynamics},
volume = {21},
number = {01n02},
pages = {1440001},
year = {2014},
doi = {10.1142/S1230161214400010}
}

@article{Banaszek1998,
  title = {Nonlocality of the Einstein-Podolsky-Rosen state in the Wigner representation},
  author = {Banaszek, Konrad and W\'odkiewicz, Krzysztof},
  journal = {Phys. Rev. A},
  volume = {58},
  issue = {6},
  pages = {4345--4347},
  numpages = {0},
  year = {1998},
  month = {Dec},
  publisher = {American Physical Society},
  doi = {10.1103/PhysRevA.58.4345},
  url = {https://link.aps.org/doi/10.1103/PhysRevA.58.4345}
}

@article{giustina2013,
  author = {Giustina, Marissa and Mech, Alexander and Ramelow, Sven and Wittmann, Björn and Kofler, Johannes and Beyer, Johannes and Lita, Adriana E. and Calkins, Brian and Gerrits, Thomas and Woo Nam, Sae and Ursin, Rupert and Wittmann, Christoph and Zeilinger, Anton},
  title = {Bell violation using entangled photons without the fair-sampling assumption},
  journal = {Nature},
  volume = {497},
  pages = {227--230},
  year = {2013},
  doi = {10.1038/nature12012},
  url = {https://doi.org/10.1038/nature12012}
}

@article{Shalm2015,
  title = {Strong Loophole-Free Test of Local Realism},
  author = {Shalm, Lynden K. and Meyer-Scott, Evan and Christensen, Bradley G. and Bierhorst, Peter and Wayne, Michael A. and Stevens, Martin J. and Gerrits, Thomas and Glancy, Scott and Hamel, Deny R. and Allman, Michael S. and Coakley, Kevin J. and Dyer, Shellee D. and Hodge, Carson and Lita, Adriana E. and Verma, Varun B. and Lambrocco, Camilla and Tortorici, Edward and Migdall, Alan L. and Zhang, Yanbao and Kumor, Daniel R. and Farr, William H. and Marsili, Francesco and Shaw, Matthew D. and Stern, Jeffrey A. and Abell\'an, Carlos and Amaya, Waldimar and Pruneri, Valerio and Jennewein, Thomas and Mitchell, Morgan W. and Kwiat, Paul G. and Bienfang, Joshua C. and Mirin, Richard P. and Knill, Emanuel and Nam, Sae Woo},
  journal = {Phys. Rev. Lett.},
  volume = {115},
  issue = {25},
  pages = {250402},
  numpages = {10},
  year = {2015},
  month = {Dec},
  publisher = {American Physical Society},
  doi = {10.1103/PhysRevLett.115.250402},
  url = {https://link.aps.org/doi/10.1103/PhysRevLett.115.250402}
}

@article{Hensen2015,
  author = {Hensen, B. and Bernien, H. and Dréau, A. E. and Reiserer, A. and Kalb, N. and Blok, M. S. and Ruitenberg, J. and Vermeulen, R. F. L. and Schouten, R. N. and Abellán, C. and Amaya, W. and Pruneri, V. and Mitchell, M. W. and Markham, M. and Twitchen, D. J. and Elkouss, D. and Wehner, S. and Taminiau, T. H. and Hanson, R.},
  title = {Loophole-free Bell inequality violation using electron spins separated by 1.3 kilometres},
  journal = {Nature},
  volume = {526},
  number = {7575},
  pages = {682--686},
  year = {2015},
  doi = {10.1038/nature15759},
  url = {https://doi.org/10.1038/nature15759},
  issn = {1476-4687}
}

@article{Zhong2019,
  title = {Violating Bell’s inequality with remotely connected superconducting qubits},
  author = {Zhong, Y. P. and Chang, H. S. and Satzinger, K. J. and Pechal, M. and Bienfait, A. and Chou, M. H. and Cleland, A. N. and Freedman, M. and Pappas, D. P. and Frunzio, L. and Devoret, M. H. and Cleland, A. N.},
  journal = {Nature Physics},
  volume = {15},
  pages = {741--744},
  year = {2019},
  doi = {10.1038/s41567-019-0507-7},
  url = {https://doi.org/10.1038/s41567-019-0507-7}
}

@article{Shin2019,
  author       = {Dong-Keun Shin and Ben M. Henson and Sean S. Hodgman and Andrew G. Truscott},
  title        = {Bell correlations between spatially separated pairs of atoms},
  journal      = {Nature Communications},
  volume       = {10},
  number       = {1},
  pages        = {4447},
  year         = {2019},
  publisher    = {Nature Publishing Group},
  doi          = {10.1038/s41467-019-12192-8},
  url          = {https://doi.org/10.1038/s41467-019-12192-8}
}

@article{Tavakoli2022,
  author       = {Armin Tavakoli and Alejandro Pozas-Kerstjens and Ming-Xing Luo and Marc-Olivier Renou},
  title        = {Bell nonlocality in networks},
  journal      = {Reports on Progress in Physics},
  volume       = {85},
  number       = {5},
  pages        = {056001},
  year         = {2022},
  publisher    = {IOP Publishing},
  doi          = {10.1088/1361-6633/ac41bb},
  url          = {https://doi.org/10.1088/1361-6633/ac41bb}
}

@article{Nongaussian_Waleschaers,
  title = {Non-Gaussian Quantum States and Where to Find Them},
  author = {Walschaers, Mattia},
  journal = {PRX Quantum},
  volume = {2},
  issue = {3},
  pages = {030204},
  numpages = {68},
  year = {2021},
  month = {Sep},
  publisher = {American Physical Society},
  doi = {10.1103/PRXQuantum.2.030204},
  url = {https://link.aps.org/doi/10.1103/PRXQuantum.2.030204}
}

@article{Wigner,
  title = {On the Quantum Correction For Thermodynamic Equilibrium},
  author = {Wigner, E.},
  journal = {Phys. Rev.},
  volume = {40},
  issue = {5},
  pages = {749--759},
  numpages = {0},
  year = {1932},
  month = {Jun},
  publisher = {American Physical Society},
  doi = {10.1103/PhysRev.40.749},
  url = {https://link.aps.org/doi/10.1103/PhysRev.40.749}
}

@article{parity_expt,
  title = {Experimental Violation of Bell's Inequality in Spatial-Parity Space},
  author = {Yarnall, Timothy and Abouraddy, Ayman F. and Saleh, Bahaa E. A. and Teich, Malvin C.},
  journal = {Phys. Rev. Lett.},
  volume = {99},
  issue = {17},
  pages = {170408},
  numpages = {4},
  year = {2007},
  month = {Oct},
  publisher = {American Physical Society},
  doi = {10.1103/PhysRevLett.99.170408},
  url = {https://link.aps.org/doi/10.1103/PhysRevLett.99.170408}
}

@article{dichotomic_nonlocality,
  title = {Quantum nonlocality test for continuous-variable states with dichotomic observables},
  author = {Jeong, H. and Son, W. and Kim, M. S. and Ahn, D. and Brukner, \ifmmode \check{C}\else \v{C}\fi{}.},
  journal = {Phys. Rev. A},
  volume = {67},
  issue = {1},
  pages = {012106},
  numpages = {7},
  year = {2003},
  month = {Jan},
  publisher = {American Physical Society},
  doi = {10.1103/PhysRevA.67.012106},
  url = {https://link.aps.org/doi/10.1103/PhysRevA.67.012106}
}

@article{Bennett1996,
  author    = {C. H. Bennett and D. P. DiVincenzo and J. A. Smolin and W. K. Wootters},
  title     = {Mixed-state entanglement and quantum error correction},
  journal   = {Phys. Rev. A},
  volume    = {54},
  pages     = {3824--3851},
  year      = {1996},
  doi       = {10.1103/PhysRevA.54.3824}
}

@article{Rains1999a,
  author    = {E. M. Rains},
  title     = {Rigorous treatment of distillable entanglement},
  journal   = {Phys. Rev. A},
  volume    = {60},
  pages     = {173--178},
  year      = {1999},
  doi       = {10.1103/PhysRevA.60.173}
}

@article{Rains1999b,
  author    = {E. M. Rains},
  title     = {Bound on distillable entanglement},
  journal   = {Phys. Rev. A},
  volume    = {60},
  pages     = {179--184},
  year      = {1999},
  doi       = {10.1103/PhysRevA.60.179}
}

@article{Eisert2002,
  author    = {J. Eisert and S. Scheel and M. B. Plenio},
  title     = {Distilling Gaussian states with Gaussian operations is impossible},
  journal   = {Phys. Rev. Lett.},
  volume    = {89},
  pages     = {137903},
  year      = {2002},
  doi       = {10.1103/PhysRevLett.89.137903}
}

@article{Giedke2002,
  author    = {G. Giedke and J. I. Cirac},
  title     = {Characterization of Gaussian operations and distillation of Gaussian states},
  journal   = {Phys. Rev. A},
  volume    = {66},
  pages     = {032316},
  year      = {2002},
  doi       = {10.1103/PhysRevA.66.032316}
}

@article{Braunstein1998,
  author    = {S. L. Braunstein},
  title     = {Error correction for continuous quantum variables},
  journal   = {Phys. Rev. Lett.},
  volume    = {80},
  pages     = {4084--4087},
  year      = {1998},
  doi       = {10.1103/PhysRevLett.80.4084}
}

@article{Navararrete2012,
  title = {Enhancing quantum entanglement by photon addition and subtraction},
  author = {Navarrete-Benlloch, Carlos and Garc\'{\i}a-Patr\'on, Ra\'ul and Shapiro, Jeffrey H. and Cerf, Nicolas J.},
  journal = {Phys. Rev. A},
  volume = {86},
  issue = {1},
  pages = {012328},
  numpages = {9},
  year = {2012},
  month = {Jul},
  publisher = {American Physical Society},
  doi = {10.1103/PhysRevA.86.012328},
  url = {https://link.aps.org/doi/10.1103/PhysRevA.86.012328}
}

@article{lee2011,
  title = {Enhancing quantum entanglement for continuous variables by a coherent superposition of photon subtraction and addition},
  author = {Lee, Su-Yong and Ji, Se-Wan and Kim, Ho-Joon and Nha, Hyunchul},
  journal = {Phys. Rev. A},
  volume = {84},
  issue = {1},
  pages = {012302},
  numpages = {6},
  year = {2011},
  month = {Jul},
  publisher = {American Physical Society},
  doi = {10.1103/PhysRevA.84.012302},
  url = {https://link.aps.org/doi/10.1103/PhysRevA.84.012302}
}

@article{Zavatta2004,
  author    = {A. Zavatta and S. Viciani and M. Bellini},
  title     = {Quantum-to-Classical Transition with Single-Photon-Added Coherent States of Light},
  journal   = {Science},
  volume    = {306},
  number    = {5696},
  pages     = {660--662},
  year      = {2004},
  doi       = {10.1126/science.1103190}
}

@article{Parigi2007,
  author    = {V. Parigi and A. Zavatta and M. S. Kim and M. Bellini},
  title     = {Probing Quantum Commutation Rules by Addition and Subtraction of Single Photons to/from a Light Field},
  journal   = {Science},
  volume    = {317},
  number    = {5846},
  pages     = {1890--1893},
  year      = {2007},
  doi       = {10.1126/science.1146204}
}

@article{boson_sampling,
  title = {Sampling arbitrary photon-added or photon-subtracted squeezed states is in the same complexity class as boson sampling},
  author = {Olson, Jonathan P. and Seshadreesan, Kaushik P. and Motes, Keith R. and Rohde, Peter P. and Dowling, Jonathan P.},
  journal = {Phys. Rev. A},
  volume = {91},
  issue = {2},
  pages = {022317},
  numpages = {6},
  year = {2015},
  month = {Feb},
  publisher = {American Physical Society},
  doi = {10.1103/PhysRevA.91.022317},
  url = {https://link.aps.org/doi/10.1103/PhysRevA.91.022317}
}

@article{Morin2014,
  author    = {O. Morin and K. Huang and J. Liu and H. Le Jeannic and C. Fabre and J. Laurat},
  title     = {Remote creation of hybrid entanglement between particle-like and wave-like optical qubits},
  journal   = {Nature Photonics},
  volume    = {8},
  pages     = {570--574},
  year      = {2014},
  doi       = {10.1038/nphoton.2014.137}
}

@book{agarwal_book,
  title     = {Quantum Optics},
  author    = {Agarwal, Girish S.},
  year      = {2012},
  publisher = {Cambridge University Press},
  address   = {Cambridge, UK}
}

@book{gerry_book,
  title     = {Introductory Quantum Optics},
  author    = {Gerry, Christopher C. and Knight, Peter L.},
  year      = {2023},
  publisher = {Cambridge University Press},
  address   = {Cambridge, UK}
}

@article{leibfried1996experimental,
  title     = {Experimental determination of the motional quantum state of a trapped atom},
  author    = {Leibfried, Dietrich and Meekhof, D. M. and King, B. E. and Monroe, C. H. and Itano, Wayne M. and Wineland, David J.},
  journal   = {Physical Review Letters},
  volume    = {77},
  number    = {21},
  pages     = {4281--4285},
  year      = {1996},
  publisher = {American Physical Society},
  doi       = {10.1103/PhysRevLett.77.4281}
}

@article{smithey1993measurement,
  title     = {Measurement of the Wigner distribution and the density matrix of a light mode using optical homodyne tomography: Application to squeezed states and the vacuum},
  author    = {Smithey, D. T. and Beck, M. and Raymer, Michael G. and Faridani, A.},
  journal   = {Physical Review Letters},
  volume    = {70},
  number    = {9},
  pages     = {1244--1247},
  year      = {1993},
  publisher = {American Physical Society},
  doi       = {10.1103/PhysRevLett.70.1244}
}

@article{Dorantes2009,
  title={Generalizations of the pseudospin operator to test the Bell inequality for the TMSV state},
  author={Dorantes, M. M. and Lucio M, J. L.},
  journal={Journal of Physics A: Mathematical and Theoretical},
  volume={42},
  number={28},
  pages={285309},
  year={2009},
  publisher={IOP Publishing},
  doi={10.1088/1751-8113/42/28/285309}
}

@article{zhang2011,
title = {Quantum nonlocality for two-mode correlated states based on generalized pseudospin operators},
journal = {Physics Letters A},
volume = {375},
number = {17},
pages = {1765-1768},
year = {2011},
issn = {0375-9601},
doi = {https://doi.org/10.1016/j.physleta.2011.03.029},
url = {https://www.sciencedirect.com/science/article/pii/S0375960111003446},
author = {Bin Zhang and Zhi-Rong Zhong and Shi-Biao Zheng}
}

@article{Kim2008,
  author       = {M. S. Kim},
  title        = {Recent developments in photon-level operations on travelling light fields},
  journal      = {Journal of Physics B: Atomic, Molecular and Optical Physics},
  volume       = {41},
  number       = {13},
  pages        = {133001},
  year         = {2008},
  publisher    = {IOP Publishing},
  doi          = {10.1088/0953-4075/41/13/133001},
  url          = {https://doi.org/10.1088/0953-4075/41/13/133001}
}

@article{mallick2025,
  title = {Efficient detection of nonclassicality using moments of the Wigner function},
  author = {Mallick, Bivas and Chakrabarty, Sudip and Mukherjee, Saheli and Maity, Ananda G. and Majumdar, A. S.},
  journal = {Phys. Rev. A},
  volume = {111},
  issue = {3},
  pages = {032406},
  numpages = {9},
  year = {2025},
  month = {Mar},
  publisher = {American Physical Society},
  doi = {10.1103/PhysRevA.111.032406},
  url = {https://link.aps.org/doi/10.1103/PhysRevA.111.032406}
}

@article{chowdhury2014,
  title = {Einstein-Podolsky-Rosen steering using quantum correlations in non-Gaussian entangled states},
  author = {Chowdhury, Priyanka and Pramanik, Tanumoy and Majumdar, A. S. and Agarwal, G. S.},
  journal = {Phys. Rev. A},
  volume = {89},
  issue = {1},
  pages = {012104},
  numpages = {10},
  year = {2014},
  month = {Jan},
  publisher = {American Physical Society},
  doi = {10.1103/PhysRevA.89.012104},
  url = {https://link.aps.org/doi/10.1103/PhysRevA.89.012104}
}

@article{Sanders_1992_ECS,
  title = {Entangled coherent states},
  author = {Sanders, Barry C.},
  journal = {Phys. Rev. A},
  volume = {45},
  issue = {9},
  pages = {6811--6815},
  numpages = {0},
  year = {1992},
  month = {May},
  publisher = {American Physical Society},
  doi = {10.1103/PhysRevA.45.6811},
  url = {https://link.aps.org/doi/10.1103/PhysRevA.45.6811}
}

@article{Van_Enk_ECS,
  title = {Entangled coherent states: Teleportation and decoherence},
  author = {van Enk, S. J. and Hirota, O.},
  journal = {Phys. Rev. A},
  volume = {64},
  issue = {2},
  pages = {022313},
  numpages = {6},
  year = {2001},
  month = {Jul},
  publisher = {American Physical Society},
  doi = {10.1103/PhysRevA.64.022313},
  url = {https://link.aps.org/doi/10.1103/PhysRevA.64.022313}
}

@article{DeFabritiis2023_ECS,
  title = {Entangled coherent states and violations of Bell-CHSH inequalities},
  author = {De Fabritiis, Philipe and Guedes, Fillipe M. and Peruzzo, Giovani and Sorella, Silvio P.},
  journal = {Physics Letters A},
  volume = {486},
  pages = {129111},
  year = {2023},
  issn = {0375-9601},
  doi = {10.1016/j.physleta.2023.129111},
  url = {https://doi.org/10.1016/j.physleta.2023.129111},
  publisher = {Elsevier}
}

@article{werner_1989,
  title = {Quantum states with Einstein-Podolsky-Rosen correlations admitting a hidden-variable model},
  author = {Werner, Reinhard F.},
  journal = {Phys. Rev. A},
  volume = {40},
  issue = {8},
  pages = {4277--4281},
  numpages = {0},
  year = {1989},
  month = {Oct},
  publisher = {American Physical Society},
  doi = {10.1103/PhysRevA.40.4277},
  url = {https://link.aps.org/doi/10.1103/PhysRevA.40.4277}
}

@article{Branciard_2010,
  title = {Characterizing the Nonlocal Correlations Created via Entanglement Swapping},
  author = {Branciard, C. and Gisin, N. and Pironio, S.},
  journal = {Phys. Rev. Lett.},
  volume = {104},
  issue = {17},
  pages = {170401},
  numpages = {4},
  year = {2010},
  month = {Apr},
  publisher = {American Physical Society},
  doi = {10.1103/PhysRevLett.104.170401},
  url = {https://link.aps.org/doi/10.1103/PhysRevLett.104.170401}
}

@article{Barnciard_2012,
  title = {Bilocal versus nonbilocal correlations in entanglement-swapping experiments},
  author = {Branciard, Cyril and Rosset, Denis and Gisin, Nicolas and Pironio, Stefano},
  journal = {Phys. Rev. A},
  volume = {85},
  issue = {3},
  pages = {032119},
  numpages = {21},
  year = {2012},
  month = {Mar},
  publisher = {American Physical Society},
  doi = {10.1103/PhysRevA.85.032119},
  url = {https://link.aps.org/doi/10.1103/PhysRevA.85.032119}
}

@article{Bartlett2002,
  title = {Efficient Classical Simulation of Continuous Variable Quantum Information Processes},
  author = {Bartlett, Stephen D. and Sanders, Barry C. and Braunstein, Samuel L. and Nemoto, Kae},
  journal = {Phys. Rev. Lett.},
  volume = {88},
  issue = {9},
  pages = {097904},
  numpages = {4},
  year = {2002},
  month = {Feb},
  publisher = {American Physical Society},
  doi = {10.1103/PhysRevLett.88.097904},
  url = {https://link.aps.org/doi/10.1103/PhysRevLett.88.097904}
}

@article{Lloyd1999,
  title = {Quantum Computation over Continuous Variables},
  author = {Lloyd, Seth and Braunstein, Samuel L.},
  journal = {Phys. Rev. Lett.},
  volume = {82},
  issue = {8},
  pages = {1784--1787},
  numpages = {0},
  year = {1999},
  month = {Feb},
  publisher = {American Physical Society},
  doi = {10.1103/PhysRevLett.82.1784},
  url = {https://link.aps.org/doi/10.1103/PhysRevLett.82.1784}
}

@article{goswami2018,
  title = {One-sided device-independent self-testing of any pure two-qubit entangled state},
  author = {Goswami, Suchetana and Bhattacharya, Bihalan and Das, Debarshi and Sasmal, Souradeep and Jebaratnam, C. and Majumdar, A. S.},
  journal = {Phys. Rev. A},
  volume = {98},
  issue = {2},
  pages = {022311},
  numpages = {7},
  year = {2018},
  month = {Aug},
  publisher = {American Physical Society},
  doi = {10.1103/PhysRevA.98.022311},
  url = {https://link.aps.org/doi/10.1103/PhysRevA.98.022311}
}

@article{Bian2020,
  title = {Experimental demonstration of one-sided device-independent self-testing of any pure two-qubit entangled state},
  author = {Bian, Zhihao and Majumdar, A. S. and Jebarathinam, C. and Wang, Kunkun and Xiao, Lei and Zhan, Xiang and Zhang, Yongsheng and Xue, Peng},
  journal = {Phys. Rev. A},
  volume = {101},
  issue = {2},
  pages = {020301},
  numpages = {6},
  year = {2020},
  month = {Feb},
  publisher = {American Physical Society},
  doi = {10.1103/PhysRevA.101.020301},
  url = {https://link.aps.org/doi/10.1103/PhysRevA.101.020301}
}

@article{adhikari2008,
  title = {Broadcasting of continuous-variable entanglement},
  author = {Adhikari, Satyabrata and Majumdar, A. S. and Nayak, N.},
  journal = {Phys. Rev. A},
  volume = {77},
  issue = {4},
  pages = {042301},
  numpages = {5},
  year = {2008},
  month = {Apr},
  publisher = {American Physical Society},
  doi = {10.1103/PhysRevA.77.042301},
  url = {https://link.aps.org/doi/10.1103/PhysRevA.77.042301}
}

@article{chowdhury2013,
  title = {Nonlocal continuous-variable correlations and violation of Bell's inequality for light beams with topological singularities},
  author = {Chowdhury, Priyanka and Majumdar, A. S. and Agarwal, G. S.},
  journal = {Phys. Rev. A},
  volume = {88},
  issue = {1},
  pages = {013830},
  numpages = {6},
  year = {2013},
  month = {Jul},
  publisher = {American Physical Society},
  doi = {10.1103/PhysRevA.88.013830},
  url = {https://link.aps.org/doi/10.1103/PhysRevA.88.013830}
}

@article{chowdhury2015,
  title = {Stronger steerability criterion for more uncertain continuous-variable systems},
  author = {Chowdhury, Priyanka and Pramanik, Tanumoy and Majumdar, A. S.},
  journal = {Phys. Rev. A},
  volume = {92},
  issue = {4},
  pages = {042317},
  numpages = {5},
  year = {2015},
  month = {Oct},
  publisher = {American Physical Society},
  doi = {10.1103/PhysRevA.92.042317},
  url = {https://link.aps.org/doi/10.1103/PhysRevA.92.042317}
}

@article{gupta2018,
  title = {Preservation of quantum nonbilocal correlations in noisy entanglement-swapping experiments using weak measurements},
  author = {Gupta, Shashank and Datta, Shounak and Majumdar, A. S.},
  journal = {Phys. Rev. A},
  volume = {98},
  issue = {4},
  pages = {042322},
  numpages = {12},
  year = {2018},
  month = {Oct},
  publisher = {American Physical Society},
  doi = {10.1103/PhysRevA.98.042322},
  url = {https://link.aps.org/doi/10.1103/PhysRevA.98.042322}
}

@article{nandi2024,
  title = {Magnetically induced Schr\"odinger cat states: The shadow of a quantum space},
  author = {Nandi, Partha and Debnath, Nandita and Kala, Subhajit and Majumdar, A. S.},
  journal = {Phys. Rev. A},
  volume = {110},
  issue = {3},
  pages = {032204},
  numpages = {15},
  year = {2024},
  month = {Sep},
  publisher = {American Physical Society},
  doi = {10.1103/PhysRevA.110.032204},
  url = {https://link.aps.org/doi/10.1103/PhysRevA.110.032204}
}

@article{Hoshi2025,
  author       = {Hoshi, D. and Nagase, T. and Kwon, S. and others},
  title        = {Entangling Schrödinger’s cat states by bridging discrete- and continuous-variable encoding},
  journal      = {Nature Communications},
  volume       = {16},
  pages        = {1309},
  year         = {2025},
  publisher    = {Nature Publishing Group},
  doi          = {10.1038/s41467-025-56503-8},
  url          = {https://doi.org/10.1038/s41467-025-56503-8}
}

@article{Tavakoli_star,
  title = {Nonlocal correlations in the star-network configuration},
  author = {Tavakoli, Armin and Skrzypczyk, Paul and Cavalcanti, Daniel and Ac\'{\i}n, Antonio},
  journal = {Phys. Rev. A},
  volume = {90},
  issue = {6},
  pages = {062109},
  numpages = {12},
  year = {2014},
  month = {Dec},
  publisher = {American Physical Society},
  doi = {10.1103/PhysRevA.90.062109},
  url = {https://link.aps.org/doi/10.1103/PhysRevA.90.062109}
}

@article{Andreoli_2017,
doi = {10.1088/1367-2630/aa8b9b},
url = {https://doi.org/10.1088/1367-2630/aa8b9b},
year = {2017},
month = {nov},
publisher = {IOP Publishing},
volume = {19},
number = {11},
pages = {113020},
author = {Andreoli, Francesco and Carvacho, Gonzalo and Santodonato, Luca and Chaves, Rafael and Sciarrino, Fabio},
title = {Maximal qubit violation of n-locality inequalities in a star-shaped quantum network},
journal = {New Journal of Physics},
}

@article{spatial_parity,
  title = {Violation of Bell's inequality with continuous spatial variables},
  author = {Abouraddy, Ayman F. and Yarnall, Timothy and Saleh, Bahaa E. A. and Teich, Malvin C.},
  journal = {Phys. Rev. A},
  volume = {75},
  issue = {5},
  pages = {052114},
  numpages = {14},
  year = {2007},
  month = {May},
  publisher = {American Physical Society},
  doi = {10.1103/PhysRevA.75.052114},
  url = {https://link.aps.org/doi/10.1103/PhysRevA.75.052114}
}

@article{Kagalwala2017,
  author  = {Kagalwala, Kumel H. and Di Giuseppe, Giovanni and Abouraddy, Ayman F. and Saleh, Bahaa E. A.},
  title   = {Single-photon three-qubit quantum logic using spatial light modulators},
  journal = {Nature Communications},
  year    = {2017},
  volume  = {8},
  number  = {1},
  pages   = {739},
  doi     = {10.1038/s41467-017-00580-x},
  url     = {https://doi.org/10.1038/s41467-017-00580-x},
}

@article{adhikari2009,
  title = {Quantum entanglement in a noncommutative system},
  author = {Adhikari, S. and Chakraborty, B. and Majumdar, A. S. and Vaidya, S.},
  journal = {Phys. Rev. A},
  volume = {79},
  issue = {4},
  pages = {042109},
  numpages = {18},
  year = {2009},
  month = {Apr},
  publisher = {American Physical Society},
  doi = {10.1103/PhysRevA.79.042109},
  url = {https://link.aps.org/doi/10.1103/PhysRevA.79.042109}
}

@article{maity2017,
  title = {Tighter Einstein-Podolsky-Rosen steering inequality based on the sum-uncertainty relation},
  author = {Maity, Ananda G. and Datta, Shounak and Majumdar, A. S.},
  journal = {Phys. Rev. A},
  volume = {96},
  issue = {5},
  pages = {052326},
  numpages = {7},
  year = {2017},
  month = {Nov},
  publisher = {American Physical Society},
  doi = {10.1103/PhysRevA.96.052326},
  url = {https://link.aps.org/doi/10.1103/PhysRevA.96.052326}
}

@article{Kimble2008,
  author  = {Kimble, H. J.},
  title   = {The quantum internet},
  journal = {Nature},
  year    = {2008},
  volume  = {453},
  number  = {7198},
  pages   = {1023--1030},
  doi     = {10.1038/nature07127},
  url     = {https://doi.org/10.1038/nature07127},
}

@mastersthesis{GuerraCopete2021,
  author  = {Guerra Copete, José Carlos},
  title   = {Network Nonlocality with Continuous Variables},
  school  = {ICFO -- Institut de Ciències Fotòniques},
  address = {Barcelona, Spain},
  year    = {2021},
  month   = aug,
  url     = {https://hdl.handle.net/2117/369248}
}

@article{Jiang2025,
  author  = {Jiang, Jun-Li and Liu, Xin-Zhu and Yang, Xue and Ding, Xiuyong and Zhang, Da and Luo, Ming-Xing},
  title   = {One-Way Network Nonlocality of Continuous Variable Entangled Networks},
  journal = {Advanced Quantum Technologies},
  year    = {2025},
  volume  = {8},
  number  = {8},
  pages   = {e2500147},
  doi     = {10.1002/qute.202500147},
  url     = {https://advanced.onlinelibrary.wiley.com/doi/10.1002/qute.202500147}
}

@article{Yarnall2007,
  title = {Synthesis and Analysis of Entangled Photonic Qubits in Spatial-Parity Space},
  author = {Yarnall, Timothy and Abouraddy, Ayman F. and Saleh, Bahaa E. A. and Teich, Malvin C.},
  journal = {Phys. Rev. Lett.},
  volume = {99},
  issue = {25},
  pages = {250502},
  numpages = {4},
  year = {2007},
  month = {Dec},
  publisher = {American Physical Society},
  doi = {10.1103/PhysRevLett.99.250502},
  url = {https://link.aps.org/doi/10.1103/PhysRevLett.99.250502}
}

@misc{kubala2026,
      title={Advanced Quantum Communication and Quantum Networks -- From basic research to future applications}, 
      author={Björn Kubala and Alexander Sauer and Alessandro Tarantola and David Fabian and Anke Ginter and Olga Kulikovska and Fabio Di Pumpo and Johannes Seiler and Wolfgang P. Schleich and Matthias Zimmermann},
      year={2026},
      eprint={2602.05781},
      archivePrefix={arXiv},
      primaryClass={quant-ph},
      url={https://arxiv.org/abs/2602.05781}, 
}

\end{document}